\documentclass{aa}  
\usepackage{courier,graphicx,epsfig,color,colordvi,natbib,url,twoopt,array,multirow}
\usepackage[varg]{txfonts}
\pdfoutput=1
\usepackage{hyperref} 
\usepackage{pdfcomment,acronym}       

\bibpunct{(}{)}{;}{a}{}{,}   

\def\# #1\par{\par\mbox{}\\ \noindent{\color{red}\small $\sharp$ #1}\\} 

\bibpunct{(}{)}{;}{a}{}{,}    
\makeatletter
 \newcommandtwoopt{\citeads}[3][][]{%
   \nonstopmode
   \href{http://adsabs.harvard.edu/abs/#3}%
        {\def\hyper@linkstart##1##2{}%
         \let\hyper@linkend\@empty\citealp[#1][#2]{#3}}
   \biblink{#3}{\href{http://adsabs.harvard.edu/abs/#3}{ADS}}%
   \errorstopmode}            
 \newcommandtwoopt{\citepads}[3][][]{%
   \nonstopmode
   \href{http://adsabs.harvard.edu/abs/#3}%
        {\def\hyper@linkstart##1##2{}%
         \let\hyper@linkend\@empty\citep[#1][#2]{#3}}
   \biblink{#3}{\href{http://adsabs.harvard.edu/abs/#3}{ADS}}%
   \errorstopmode}            
 \newcommandtwoopt{\citetads}[3][][]{%
   \nonstopmode
   \href{http://adsabs.harvard.edu/abs/#3}%
        {\def\hyper@linkstart##1##2{}%
         \let\hyper@linkend\@empty\citet[#1][#2]{#3}}
   \biblink{#3}{\href{http://adsabs.harvard.edu/abs/#3}{ADS}}%
   \errorstopmode}            
 \newcommandtwoopt{\citeyearads}[3][][]{%
   \nonstopmode
   \href{http://adsabs.harvard.edu/abs/#3}%
        {\def\hyper@linkstart##1##2{}%
         \let\hyper@linkend\@empty\citeyear[#1][#2]{#3}}
   \biblink{#3}{\href{http://adsabs.harvard.edu/abs/#3}{ADS}}%
   \errorstopmode}            
\makeatother

\newcount\longrefs
\def\aap{\ifnum\longrefs=1 {Astron.\ Astrophys.}\else 
                           {A\hbox{\rm \&}A}\fi}
\def\aapr{\ifnum\longrefs=1 {Astron.\ Astrophys.\ Rev.}\else 
                            {A\hbox{\rm \&}AR}\fi}
\def\aaps{\ifnum\longrefs=1 {Astron.\ Astrophys.\ Suppl.}\else 
                            {A\hbox{\rm \&}A Suppl.}\fi}
\def\actaa{\ifnum\longrefs=1 {Acta Astronomica}\else
                            {Acta Astron.}\fi}
\def\aipcs{\ifnum\longrefs=1 {Am.\ Inst.\ Phys.\ Conf.\ Series}\else
                             {AIP Conf.\ Ser.}\fi}
\def\aj{\ifnum\longrefs=1 {Astron.\ J.}\else 
                          {AJ}\fi} 
\def\ao{\ifnum\longrefs=1 {Applied Optics}\else 
                           {Appl.\ Opt.}\fi} 
\def\aspcs{\ifnum\longrefs=1 {Astron.\ Soc.\ Pacific Conf.\ Series}\else 
                           {ASP Conf.\ Ser.}\fi} 
\def\apj{\ifnum\longrefs=1 {Astrophys.\ J.}\else 
                           {ApJ}\fi} 
\def\apjl{\ifnum\longrefs=1 {Astrophys.\ J. Lett.}\else 
                            {ApJL}\fi} 
\def\aplett{\ifnum\longrefs=1 {Astrophys.\ J. Lett.}\else 
                            {ApJ}\fi} 
\def\apjs{\ifnum\longrefs=1 {Astrophys.\ J. Suppl.}\else 
                            {ApJS}\fi}
\def\apss{\ifnum\longrefs=1 {Astrophys.\ and Space Science}\else 
                            {Astrophys.\ Space Sci.}\fi}
\def\araa{\ifnum\longrefs=1 {Ann.\ Rev.\ Astron.\ Astrophys.}\else 
                            {ARA\hbox{\rm \&}A}\fi}
\def\azh{\ifnum\longrefs=1 {Astronomicheskii Zhurnal}\else 
                            {Astron.\ Zhur.}\fi}
\def\baas{\ifnum\longrefs=1 {Bull.\ Am.\ Astron.\ Soc.}\else 
                            {BAAS}\fi}
\def\bain{\ifnum\longrefs=1 {Bull.\ Astronom.\ Institutes Netherlands}\else
                            {Bull.\ Astr.\ Inst.\ Neth.}\fi}
\def\cjaa{\ifnum\longrefs=1 {Chinese Jour.\ Astron.\ Astrophys.}\else 
                            {Chin.\ J.\ A\&A}\fi}
\def\gca{\ifnum\longrefs=1 {Geochim.\ Cosmochim.\ Acta}\else 
                           {Geochim.\ Cosmochim.\ Acta}\fi}
\def\grl{\ifnum\longrefs=1 {Geophys.\ Res.\ Lett.}\else 
                           {Geoph.\ Res.\ Lett.}\fi}
\def\iaucirc{\ifnum\longrefs=1 {IAU Circulars}\else 
                          {IAU Circ.}\fi}
\def\icarus{\ifnum\longrefs=1 {Icarus}\else 
                          {Icarus}\fi}
\def\ip{\ifnum\longrefs=1 {in press}\else 
                          {in press}\fi}
\def\jcap{\ifnum\longrefs=1 {Jour.\ Cosmology Astropart.\ Phys.}\else 
                          {JCAP}\fi}
\def\jgr{\ifnum\longrefs=1 {J.\ Geophys.\ Res.}\else 
                           {J.\ Geophys.\ Res.}\fi}  
\def\jrasc{\ifnum\longrefs=1 {J.\ Royal Astron.\ Soc.\ Canada}\else 
                           {JRAS Can.}\fi}  
\def\memsai{\ifnum\longrefs=1 {Mem.~Soc.~Astron.~Italiana}\else
                              {MmSAI}\fi}
\def\mnras{\ifnum\longrefs=1 {Mon.\ Not.\ Roy.\ Astron.\ Soc.}\else 
                             {MNRAS}\fi} 
\def\na{\ifnum\longrefs=1 {New Astronomy}\else 
                           {New Astron.}\fi}
\def\nar{\ifnum\longrefs=1 {New Astronomy rev.}\else 
                           {New Astron.\ Rev.}\fi}
\def\nat{\ifnum\longrefs=1 {Nature}\else 
                           {Nat}\fi}
\def\pasa{\ifnum\longrefs=1 {Pub.\ Astron.\ Soc.\ Australia}\else 
                            {PASA}\fi} 
\def\pasj{\ifnum\longrefs=1 {Pub.\ Astron.\ Soc.\ Japan}\else 
                            {PASJ}\fi} 
\def\pasp{\ifnum\longrefs=1 {Pub.\ Astron.\ Soc.\ Pacific}\else 
                            {PASP}\fi} 
\def\physscr{\ifnum\longrefs=1 {Physica Scripta}\else 
                            {Phys.\ Scrip.}\fi} 
\def\planss{\ifnum\longrefs=1 {Planetary \& Space Science}\else 
                            {Plan. \& Space Sci.}\fi} 
\def\procspie{\ifnum\longrefs=1 {Proc.\ SPIE}\else 
                            {Proc.\ SPIE}\fi} 
\def\qjras{\ifnum\longrefs=1 {Quarterly J.\ Royal Astron.\ Soc.}\else 
                            {QJRAS}\fi} 
\def\rmxaa{\ifnum\longrefs=1 {Revista Mexicana de Astron.\ y Astrofys.}\else 
                            {RMxAA}\fi} 
\def\sa{\ifnum\longrefs=1 {Soviet Astron..}\else 
                               {Sov.\ Astron.}\fi}
\def\skytel{\ifnum\longrefs=1 {Sky \& Telescope}\else 
                            {Sky \& Tel.}\fi} 
\def\solphys{\ifnum\longrefs=1 {Solar Phys.}\else 
                               {SoPh}\fi}
\def\sovast{\ifnum\longrefs=1 {Soviet Astronomy}\else 
                               {Sov.\ Ast.}\fi}
\def\ssr{\ifnum\longrefs=1 {Space Science Rev.}\else 
                               {Space\ Sci.\ Rev.}\fi}
\def\zap{\ifnum\longrefs=1 {Zeitschr.\ f.\ Astrophysik}\else
                               {Z.\ Astrophys.}\fi}

\makeatletter
\newcommand{\bibnote}[2]{\@namedef{#1note}{#2}}
\newcommand{\biblink}[2]{\@namedef{#1link}{#2}}
\makeatother

\newacro{AA}{Astronomy \& Astrophysics}
\newacro{ADS}{Astrophysics Data System}
\newacro{AIA}{Atmospheric Imaging Assembly}
\newacro{AO}{adaptive optics}
\newacro{ApJ}{Astrophysical Journal}
\newacro{AR}{active region}
\newacro{BFI}{Broad-band Filter Imager}
\newacro{CE}{coronal equilibrium}
\newacro{CfA}{Center for Astrophysics}
\newacro{CME}{coronal mass ejection}
\newacro{CRD}{complete redistribution}
\newacro{CRISP}{CRisp Imaging SpectroPolarimeter}
\newacro{CRISPEX}{CRisp SPectral EXplorer}
\newacro{CS}{coherent scattering}
\newacro{DEM}{Differential Emission Measure}
\newacro{DF}{dynamic fibril}
\newacro{DKIST}{Daniel K. Inouye Solar Telescope}
\newacro{DLR}{Deutsches Zentrum f\"ur Luft- und Raumfahrt}
\newacro{DOT}{Dutch Open Telescope}
\newacro{DST}{Richard B. Dunn Solar Telescope}   
\newacro{EB}{Ellerman bomb}
\newacro{EDP}{\'{E}dition Diffusion Presse Sciences}  
\newacro{EIT}{Extreme ultraviolet Imaging Telescope}
\newacro{EPIC}{European participation in Solar-C}
\newacro{ERC}{European Research Council}
\newacro{ESA}{European Space Agency}
\newacro{EST}{European Solar Telescope}
\newacro{EUV}{extreme ultraviolet}
\newacro{FAF}{flaring active-region fibril}
\newacro{FITS}{Flexible Image Transport System}
\newacro{FOV}{field of view}
\newacro{fov}{field of view}
\newacro{FWHM}{full width at half maximum}
\newacro{HAO}{High Altitude Observatory}
\newacro{HD}{hydrodynamics}
\newacro{Hi-C}{High Resolution Coronal Imager Sounding Rocket}
\newacro{HMI}{Helioseismic and Magnetic Imager}
\newacro{IAA}{Instituto de Astrof\'{i}sica de Andaluc\'{i}a}
\newacro{IAC}{Instituto de Astrof\'{i}sica de Canarias}
\newacro{IAS}{Institut d'Astrophysique Spatiale}
\newacro{IDL}{Interactive Data Language}
\newacro{IMaX}{Imaging Magnetograph eXperiment}
\newacro{INAF}{Istituto Nazionale di Astrofisica}
\newacro{IB}{IRIS bomb}
\newacro{IR}{infrared}
\newacro{IRIS}{Interface Region Imaging Spectrograph}
\newacro{ISAS}{Institute of Space and Astronautical Science}
\newacro{ISP}{Institute for Solar Physics}
\newacro{ISS}{International Space Station}
\newacro{ISSI}{International Space Science Institute}
\newacro{ITA}{Institute for Theoretical Astrophysics}
\newacro{JAXA}{Japan Aerospace Exploration Agency}
\newacro{KIS}{Kiepenheuer--Institut f\"{u}r Sonnenphysik}
\newacro{KPNO}{Kitt Peak National Observatory}
\newacro{LASP}{Laboratory for Atmospheric and Space Physics}
\newacro{LC}{liquid cristal}
\newacro{LMSAL}{Lockheed Martin Solar and Astrophysics Labratory}
\newacro{LOS}{line of sight}
\newacro{LTE}{local thermodynamic equilibrium}
\newacro{MC}{magnetic concentration}
\newacro{MCAO}{multi-conjugate adaptive optics} 
\newacro{MDI}{Michelson Doppler Imager}
\newacro{ME}{Milne-Eddington} 
\newacro{MHD}{magnetohydrodynamics}
\newacro{MOMFBD}{Multi-Object Multi-Frame Blind Deconvolution}
\newacro{MPE}{Max--Planck--Institut f\"ur extraterrestrische Physik}
\newacro{MPG}{Max--Planck--Gesellschaft}
\newacro{MPS}{Max Planck Institute for Solar System Research}
\newacro{MSSL}{Mullard Space Science Laboratory}
\newacro{MTF}{modulation transfer function}
\newacro{NAOJ}{National Astronomical Observatory of Japan}
\newacro{NASA}{National Aeronautics and Space Administration}
\newacro{NLTE}{non-local thermodynamic equilibrium}
\newacro{NLFFF}{non-linear force-free field}
\newacro{NOAA}{National Oceanic and Atmospheric Administration}
\newacro{non-E}{non-equilibrium}
\newacro{NSO}{National Solar Observatory}
\newacro{NWO}{Netherlands Organisation for Scientific Research}
\newacro{PRD}{partial redistribution}
\newacro{PROBA2}{PRoject for OnBoard Autonomy}
\newacro{PSF}{point spread function}
\newacro{QS}{quiet Sun}
\newacro{QSEB}{quiet-Sun Ellerman-like brightening} 
\newacro{RAL}{Rutherford Appleton Laboratory}
\newacro{RBE}{rapid blue-shifted excursion}
\newacro{R-MHD}{radiation hydrodynamics}
\newacro{rms}{root mean square}
\newacro{RMS}{root mean square}
\newacro{ROB}{Royal Observatory of Belgium}
\newacro{ROI}{region of interest}
\newacro{RRE}{rapid red-shifted excursion}
\newacro{RTE}{radiative transfer equation}
\newacro{SE}{statistical equilibrium}
\newacro{SB}{Saha Boltzmann}
\newacro{SDO}{Solar Dynamics Observatory}
\newacro{SJI}{slit-jaw image}
\newacro{SNR}{signal-to-noise ratio}
\newacro{SO}{Solar Orbiter}
\newacro{SoHO}{Solar and Heliospheric Observatory}
\newacro{SP}{Spectropolarimeter}
\newacro{SST}{Swedish 1-m Solar Telescope}
\newacro{SUMER}{Solar Ultraviolet Measurements of Emitted Radiation}
\newacro{SUFI}{Sunrise Filter Imager}
\newacro{SVD}{singular value decomposition}
\newacro{SVST}{Swedish Vacuum Solar Telescope}
\newacro{THEMIS}{T\'{e}lescope H\'{e}liographique pour l'Etude du 
   Magn\'{e}tisme et des Instabilit\'{e} Solaires}     
\newacro{TR}{transition region}
\newacro{TRACE}{Transition Region and Coronal Explorer}
\newacro{TSI}{total solar irradiance}
\newacro{UT}{Universal Time}
\newacro{UV}{ultraviolet}
\newacro{VAULT}{Very high angular resolution ultraviolet telescope}
\newacro{VIRGO}{Variability of solar IRradiance and Gravity Oscillations}
\newacro{VTT}{Vacuum Tower Telescope}    
\newacro{XRT}{X-Ray Telescope}

\long\def\startignore #1\stopignore{}   

\def\rmit#1{{\it #1}}              
           
\def\ie{\rmit{i.e.,}}              
\def\eg{\rmit{e.g.,}}              
\def\cf{cf.}                       

\def\specchar#1{\uppercase{#1}}    
\def\specand{ and }                
\def\specand{\,\&\,}               
\def\AlI{\mbox{Al\,\specchar{i}}}  

\def\CII{\mbox{C\,\specchar{ii}}} 
 
\def\CIV{\mbox{C\,\specchar{iv}}}

\def\FeI{\mbox{Fe\,\specchar{i}}}

\def\MgI{\mbox{Mg\,\specchar{i}}} 
\def\MgII{\mbox{Mg\,\specchar{ii}}}

\def\SiI{\mbox{Si\,\specchar{i}}}

\def\SiIV{\mbox{Si\,\specchar{iv}}}

\def\Halpha{\mbox{H\hspace{0.1ex}$\alpha$}} 

\def\HeIDthree{\mbox{He\,\specchar{i}\,\,D$_3$}}

\def\NaID{\mbox{Na\,\specchar{i}\,\,D}}

\def\MgIb{\mbox{Mg\,\specchar{i}\,b}}

\def\CaIIH{\mbox{Ca\,\specchar{ii}\,\,H}}

\def\CaIR{\mbox{Ca\,\specchar{ii}\,8542\,\AA}} 

\def\hk{\mbox{h{\specand}k}}

\def\level #1 #2#3#4{$#1 \; ^{#2} \mbox{#3} ^{#4}$}   

\def\deg{\hbox{$^\circ$}}       

\def\arcsec{\hbox{$^{\prime\prime}$}}

\def\={\hbox{$\!=\!$}}                     

\def\specchar#1{{\sc{#1}}}    
\def\rmit#1{#1}               

\def\deg{\hbox{$^\circ$}}      
\def\mAA{m\AA}
\def\sqasec{\hbox{arcsec$^{2}$}}
\def\Halphawav{\Halpha~6563\,\AA}
\def\HeIR{\mbox{He\,\specchar{i}\,\,10830\,\AA}}
          
\def\EB{Ellerman bomb}
\def\EBs{Ellerman bombs}
\def\UVB{UV burst}
\def\UVBs{UV bursts}
\def\SDO{{\it Solar Dynamics Observatory\/}}
\def\AIA{{\it Atmospheric Imaging Assembly\/}}
\def\HMI{{\it Helioseismic and Magnetic Imager\/}}
\def\SST{Swedish 1-m Solar Telescope}

\def\EBDETECT{{\tt EBDETECT}}

\def\IRIS{{\it Interface Region Imaging Spectrograph}}
\def\TRACE{{\it Transition Region and Coronal Explorer\/}}

\def\fone{\hbox{F$_{1}$}}

\newcommand{\fov}[2]{{{#1}\arcsec$\times${#2}\arcsec}}

\newcommand{\gregalemail}{gregal.vissers@astro.su.se}

\begin{document}

\title{Automating Ellerman bomb detection in ultraviolet continua}

\author{Gregal J. M. Vissers\inst{1,2}
\and
Luc H. M. Rouppe van der Voort\inst{2,3}
\and
Robert J. Rutten\inst{2,3,4}}
\institute{Institute for Solar Physics, Department of Astronomy, 
 Stockholm University, AlbaNova University Centre,
 106 91 Stockholm, Sweden; \gregalemail
\and
Institute of Theoretical Astrophysics,
  University of Oslo, %
  P.O. Box 1029 Blindern, N-0315 Oslo, Norway 
\and Rosseland Centre for Solar Physics, 
  University of Oslo, %
P.O. Box 1029 Blindern, N-0315 Oslo, Norway
\and
Lingezicht Astrophysics, 't Oosteneind 9, 4158\,CA Deil, The Netherlands}


\authorrunning{Vissers et al.}

\abstract{ \EBs\ are transient brightenings in the wings of
\Halphawav\ that pinpoint photospheric sites of magnetic reconnection
in solar active regions.
Their partial visibility in the 1600\,\AA\ and 1700\,\AA\ continua
registered routinely by the \AIA\ (AIA) onboard the \SDO\ (SDO) offers
a unique opportunity to inventory such magnetic-field disruptions
throughout the AIA database if a reliable recipe for their detection
can be formulated. 
This is done here.
We improve and apply an \Halpha\ \EB\ detection code to ten data sets
spanning viewing angles from solar disc centre to the limb. 
They combine high-quality \Halpha\ imaging spectroscopy from the \SST\
with simultaneous AIA imaging around 1600\,\AA\ and 1700\,\AA.
A trial grid of brightness, lifetime and area constraints is imposed
on the AIA images to define optimal recovery of the 1735 Ellerman
bombs detected in \Halpha.
The best results when optimising simultaneously for recovery fraction
and reliability are obtained from 1700\,\AA\ images by requiring
5$\sigma$\ brightening above the average 1700\,\AA\ nearby quiet-Sun
intensity, lifetime above one minute, area of 1--18 AIA pixels.
With this recipe 27\% of the AIA detections are \Halpha-detected \EBs\
while it recovers 19\% of these (of which many are smaller than the
AIA resolution).
Better yet, among the top 10\% AIA 1700\,\AA\ detections selected with
combined brightness, lifetime and area thresholds as many as 80\% are
\Halpha\ \EBs.
Automated selection of the best 1700\,\AA\ candidates therefore opens
the entire AIA database for detecting most of the more significant
photospheric reconnection events.
This proxy is applicable as flux-dynamics tell-tale in studying any
Earth-side solar active region since early 2010 up to the present.
}

\keywords{Sun: activity -- Sun: atmosphere -- Sun: magnetic fields -- Sun: UV
radiation}

\maketitle

\section{Introduction}\label{sec:introduction}
Ellerman bombs
\citepads{1917ApJ....46..298E} 
are among the most spectacular small-scale eruptive events in the
solar spectrum.
In the outer wings of \Halphawav\ they display flame morphology with
highly dynamic sub-structuring at the resolution limit of current
high-resolution telescopes (%
\eg\
\citeads{2010PASJ...62..879H}, 
\citeads{2011ApJ...736...71W}, 
\citeads{2015ApJ...798...19N}, 
\citeads{2017ApJ...851L...6R}). 
They are predominantly observed near polarity inversion lines (\eg\
\citeads{2002ApJ...575..506G}, 
\citeads{2006ApJ...643.1325F}, 
\citeads{2007A&A...473..279P}, 
\citeads{2008PASJ...60..577M}, 
\citeads{2010PASJ...62..879H}, 
\citeads{2013ApJ...774...32V}, 
\citeads{2016ApJ...823..110R}) 
in regions where magnetic field patterns on the solar surface change
much, such as emerging flux regions and rapidly emerging or decaying
active regions.
They pinpoint reconnection in the solar photosphere.

The numerical simulations of
\citetads{2009A&A...508.1469A} 
already suggested that \EBs\ represent a flux-emergence phenomenon.
More advanced numerical simulations including \Halphawav\ synthesis
have recently confirmed that small-scale magnetic reconnection within
the photosphere is indeed their driving agent.
\citetads{2017ApJ...839...22H} 
did so for stronger-field events with the Bifrost code;
\citetads{2017A&A...601A.122D} 
for weaker-field events with the MURaM code.
The latter resemble the quiet-Sun Ellerman-like brightenings (QSEB)
discovered by \citetads{2016A&A...592A.100R}. 

\EBs\ are classically identified through their large \Halphawav\ wing
brightening, but they can also be seen as wing enhancements in other
chromospheric lines including %
\CaIIH\ at 3968\,\AA\ 
(\citeads{2008PASJ...60..577M}, 
\citeads{2010PASJ...62..879H}, 
\citeads{2015A&A...582A.104R}), 
\CaIR\
(\citeads{2006ApJ...643.1325F}, 
\citeads{2006SoPh..235...75S}, 
\citeads{2007A&A...473..279P}, 
\citeads{2013ApJ...779..143R}, 
\citeads{2013ApJ...774...32V}), 
\HeIDthree\ at 5876\,\AA\ 
and \HeIR\ \citepads{2017A&A...598A..33L}, 
and in the ultraviolet sampled by the \IRIS\ (IRIS;
\citeads{2014SoPh..289.2733D}) 
as enhancements of the \MgII, \CII\ and \SiIV\ resonance lines
(\citeads{2015ApJ...812...11V}, 
\citeads{2016A&A...593A..32G}). 
Of these the \SiIV\ lines near 1400\,\AA\ are most informative because
Ellerman bombs can appear optically thin in these and display
bimodal-jet structure directly
(\citeads{2015ApJ...812...11V}). 

\EBs\ are generally also observed as brightenings in the 1600\,\AA\
and 1700\,\AA\ continua that normally originate from the upper
photosphere (\eg\
\citeads{2000ApJ...544L.157Q}, 
\citeads{2002ApJ...575..506G}, 
\citeads{2007A&A...473..279P}, 
\citeads{2010MmSAI..81..646B}, 
\citeads{2011CEAB...35..181H}, 
\citeads{2013ApJ...774...32V}, 
\citeads{2013JPhCS.440a2007R}, 
\citeads{2015A&A...582A.104R}, 
\citeads{2017GSL.....4...30C}). 
However, they have not been detected in the \NaID\ and \MgIb\ lines
(\citeads{1917ApJ....46..298E}, 
\citeads{2015ApJ...808..133R}) 
formed at similar heights.
This apparent contradiction was attributed by
\citetads{2016A&A...590A.124R} 
to ionisation of the neutral metal stages combined with
non-equilibrium over-opacity in the scattering Balmer continuum, but
this issue has not yet been addressed with numerical modelling.  
Non-equilibrium simulation and spectral synthesis are likely required
to explain it, but even without understanding the brightness
signatures of \EBs\ in mid-ultraviolet continua we may yet exploit
them for \EB\ detection and localisation.

The launch of the \SDO\ (SDO) in February 2010 has resulted in
continuous monitoring of the whole Earth-facing side of the Sun ever
since (and hopefully for years to come) in nine ultraviolet passbands including
wide ones around 1600\,\AA\ and around 1700\,\AA\ with the \AIA\
(AIA).
SDO also collects photospheric magnetograms with the \HMI\ (HMI), but
recognising small-scale reconnection events from photospheric flux
cancelations requires higher angular resolution and magnetic
sensitivity than what HMI provides.
While deep learning techniques may improve this shortcoming
\citepads{2018A&A...614A...5D}, 
\EBs\ already present a viable alternative to locate photospheric
reconnection events, not by displaying the bipolar input prior to
cancelation but by displaying the resulting energy output. 

Thus, the occurrence of \EBs\ may be used to detect and trace
small-scale solar magnetic field reorganisation, making them an effective
proxy for (on-going) flux emergence and an early warning of solar activity. 
This makes it desirable to be able to pinpoint \EBs\ anywhere at any
time, not only from highest-quality \Halpha\ observing as done so far.
The latter requires the very best seeing at the very best telescopes
and is therefore severely limited to rare observation, short sampling
duration, and small field of view so that most \EB\ studies have
analysed only one or only a few.
Characterising \EB\ signatures in mid-ultraviolet images and defining
a reliable detection recipe that only requires the ubiquitous AIA data
(over 10 million 1600--1700\,\AA\ full-disc image pairs to date) is
obviously a worthwhile quest.
We undertake that here.
A similar \Halpha--AIA 1700\,\AA\ correspondence study was recently
done by \citetads{2017GSL.....4...30C} 
but only for a single observation; here, we cover the full centre-limb
variation and a variety of active regions by analysing ten different
data sets.   

Reliable identification of \EBs\ is non-trivial both in the \Halpha\
wings and in the ultraviolet continua due to competing small-scale
brightness features of differing nature.
At low brightness these are the magnetic concentrations that
constitute network and plage and were described as ``magnetic bright
points'' and modelled as magnetostatic fluxtubes in the older
literature. 
Their observation requires sub-arcsecond resolution to avoid
cancelation of their brightness against the darkness of the
intergranular lanes in which they reside
(\citeads{1996ApJ...463..797T}) 
and only with the superior resolution of the \SST\ (SST;
\citeads{2003SPIE.4853..341S}) 
they were resolved into intricate morphologies
(\citeads{2004A&A...428..613B}). 
These concentrations also show enhanced brightening in the \Halpha\
wings (Leenaarts
\citeyearads{2006A&A...452L..15L}, 
\citeyearads{2006A&A...449.1209L}), 
due to deeper fluxtube hole visibility through reduced collisional
damping at lower density.
We call these ubiquitous brightenings ``pseudo-EBs'' following
\citetads{2013JPhCS.440a2007R} 
who noted that a significant fraction of the \EB\ literature
mistakenly addressed them, although
\citetads{1917ApJ....46..298E} 
already warned against facular confusion.

The non-reconnecting magnetic concentrations also appear bright in
ultraviolet continua, probably from larger fluxtube transparency by
ionisation of \SiI, \MgI, \FeI\ and \AlI, so that lower-brightness
selection thresholds must be used not only for \Halpha\ but also for
the 1600 and 1700\,\AA\ images. 
The 1600\,\AA\ images display \EBs\ at higher contrast over the
quiescent network than the 1700\,\AA\ images, and also larger
(Fig.~\ref{fig:aia_ebfaf} below).
These enhancements are likely due to extra emission and scattering in
the \CIV\ resonance lines in the 1600\,\AA\ passband. 

At high brightness the major cause of misidentification are the
flaring active-region fibrils (FAF) described in
\citetads{2015ApJ...812...11V} 
and \citetads{2016A&A...590A.124R} 
and possibly named microflares elsewhere.
They appear primarily in the 1600\,\AA\ images, probably due to large
\CIV\ contribution; in the 1700\,\AA\ images they are much weaker or
absent. 
In 1600\,\AA\ movies one recognises them by their sudden appearance,
large brightness, extended elongated shape, and very fast apparent
motion along filamentary tracks.   
They, or their aftermaths, show large emission in the IRIS lines and
also leave signatures in the AIA EUV passbands
(\citeads{2015ApJ...812...11V}), 
whereas \EBs\ may show up in the IRIS lines too but weaker and do not
affect the overlying fibrils observed in the core of \Halpha\ 
(as already remarked by \citeads{1917ApJ....46..298E}) 
and similarly in the cores of \MgII\ \hk.

More generally sudden, small, energetic brightenings observed in the
ultraviolet are called \UVBs.
They are reviewed comprehensively in
\citetads{2018SSRv..214..120Y} 
and include the IRIS bursts of
\citetads{2014Sci...346C.315P}, 
FAFs, and also part of the \Halpha-identified \EB\ population but
without complete overlap %
(\citeads{2015ApJ...812...11V}, 
\citeads{2015ApJ...810...38K}, 
\citeads{2016ApJ...824...96T}, 
\citeads{2017A&A...598A..33L}). 
This partial non-correspondence seems primarily due to difference in
height of the energy-releasing reconnection event.
Observationally this is suggested by common UV-burst response in
chromospheric and transition-region diagnostics whereas \EBs\ show no
counterpart in the hotter AIA channels.
Computationally it is suggested by the simulation results of
\citetads{2017ApJ...839...22H} 
who reproduced \EBs\ from lower-height reconnection and \UVBs\ from
larger-height reconnection, but this simulation did not produce both
signatures simultaneously whereas the observed populations do overlap.

The observational mix of small-feature brightenings in the form of
pseudo-EBs, bona fide \EBs, FAFs, and other \UVBs\ make unequivocal
\EB\ identification a difficult task.  
Over the past years we have gained considerable experience in
recognising them in the many pertinent \Halpha\ data sets which the
Oslo group collected at the SST, employing the versatile CRISPEX
browser (%
\citeads{2012ApJ...750...22V}, 
\citeads{2018arXiv180403030L}) 
for interactive inspection.
A major \Halpha\ tell-tale is flame morphology in limbward viewing,
first demonstrated in \citetads{2011ApJ...736...71W} 
and also the key diagnostic in recognising QSEBs
(\citeads{2016A&A...592A.100R}). 
Others are the sudden \EB\ appearance and their rapid fine-structure
variation. 
Together, these give us confidence in separating \EBs\ from network
pseudo-EBs in high quality SST data.
In 1600\,\AA\ images smallness, roundish non-filamentary shape, and
larger stationarity distinguish \EBs\ from FAFs.

Our expertise in \EB\ identification in SST data relies critically on
the superior resolution obtained with this superb telescope.
In contrast, AIA's resolution is ten times worse; we show below that
most SST \Halpha\ \EBs\ are smaller than a single AIA pixel. 
This suggests that recovering \Halpha\ \EBs\ in AIA images is hampered
severely by lack of resolution, but it should be noted that whereas
\EBs\ have optically thin formation of their outer \Halpha\ wing
brightenings, permitting scattering-free intensity variations over
very fine scales (``striations'') as indeed observed
(\citeads{2011ApJ...736...71W}), 
the ultraviolet continua are strongly scattering which results in
apparent feature spreading and smoothing over a few hundred km
(Fig.~36 of \citeads{1981ApJS...45..635V}). 
Also, the steeper Planck function sensitivity at shorter wavelengths
enhances ultraviolet contrasts.
Thus, very fine sub-pixel intensity spikes can still cause
full-AIA-pixel brightening.

\begin{figure*}[p]
  \centerline{\includegraphics[width=0.97\textwidth]{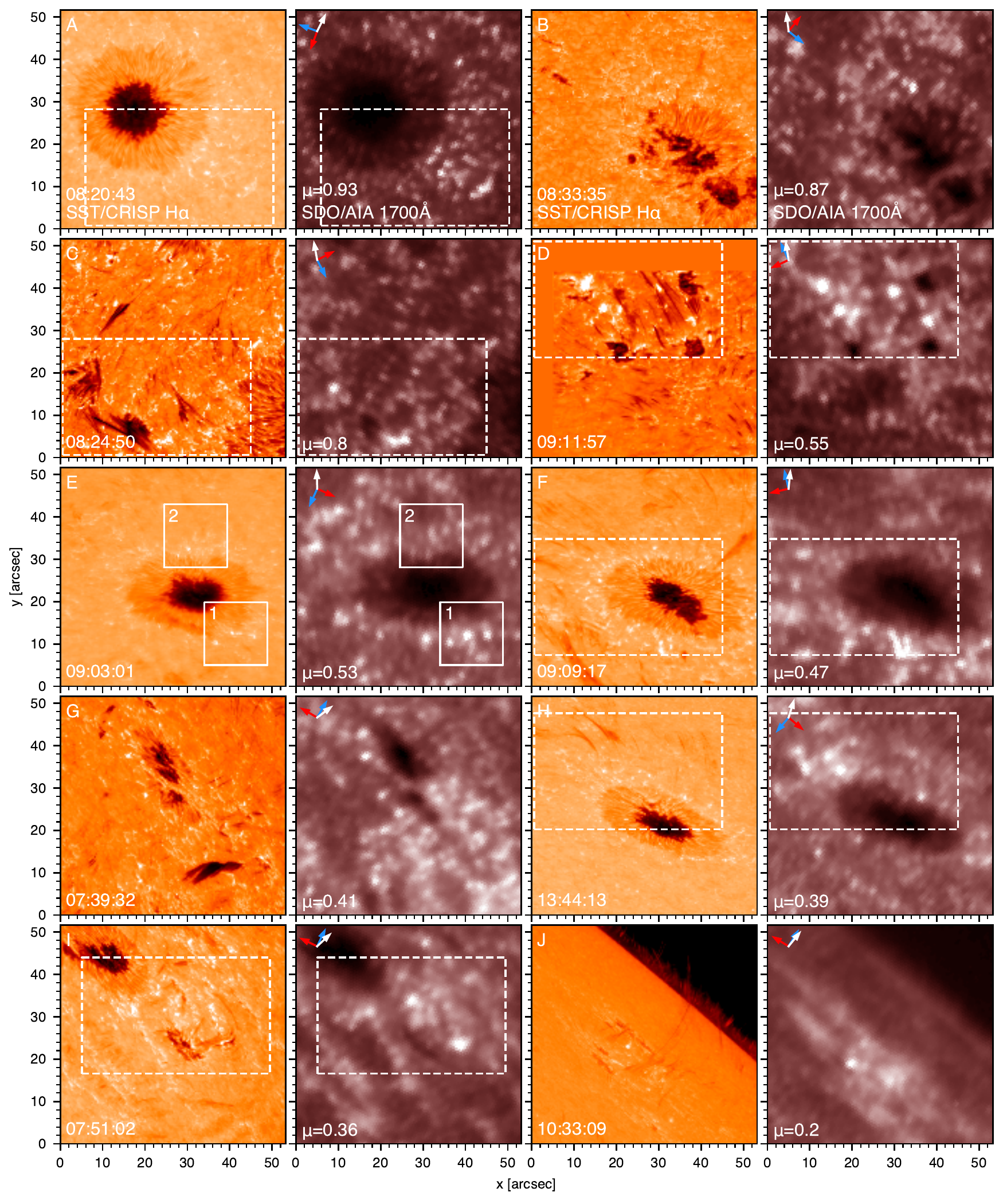}}
  \vspace{-3ex}
  \caption[]{\label{fig:fovs} %
  Full field-of-view samples of all 10 data sets.
  Panel pairs (A)--(J) show near-simultaneous co-aligned CRISP
  \Halpha-wing images (orange; blue-wing images averaged around $-$1\,\AA\ for
  all but data set B that shows the red wing at +1\,\AA) and AIA 1700\,\AA\ images
  (red-brown),
  ordered by decreasing viewing angle $\mu$ ($=\cos \theta$, with
  $\theta$ the angle between the line-of-sight and the normal to the solar
surface) 
  specified at lower-left in each AIA
  panel with the image-center solar $(X,Y)$ location.
  The times of the SST observations are specified at lower-left in
  each CRISP panel.  
  The corresponding AIA image cutouts were interpolated to these from
  their 24\,s sampling cadence (through nearest-neighbour frame selection) and rotated to the SST orientation.
  Each field of view has been cut slightly to obtain the same display
  size and scale. 
  The arrows at top left in each AIA panel point towards solar North
  ({\it red\/}), West ({\it blue\/}) and the nearest limb ({\it
  white\/}). 
  Each image is byte-scaled independently including high-level clipping
  to improve the overall scene visibility.
  Dashed frames in pairs C, D, F and I define cutouts for
  Fig.~\ref{fig:aia_ebfaf}, solid frames in pair E define cutouts for
  Fig.~\ref{fig:detection_comparison}.
  }
\end{figure*}

\begin{table*}[bth]
\caption{Overview over the data sets analysed in this study.}
\begin{center}
\begin{tabular}{clcccllccccc}%
        \hline \hline
        {} & {}          & \multicolumn{3}{c}{Target}              & {}          & \multicolumn{3}{c}{Diagnostic Details}            & {}          & {} \\
  \cline{3-4} \cline{6-9}
  {}    & {}                  & {}        & ($X$,$Y$) [\arcsec]
  & $\theta$         & Instru-          & {}      & Range & $\Delta\lambda$ & $\Delta t$  &       Time \\
        Set     & Date        & AR        & FOV size       & ($\mu$)          & ment  & Filter  & [\AA] & [\mAA]                & [s]                 & (UTC) \\
  \hline
  A   & 2010 Jul 2  & 11084     & ($-$38,$-$346)  & 21.6$\deg$        & 
        CRISP   & \Halpha & $\pm$1.9  & 85          & 22.4        & 07:30\,--\,08:42 \\
  {}  & {}          & MF        & \fov{53}{52}    & (0.93)            & 
        AIA & \multicolumn{3}{l}{1600 / 1700} & 24.0        & {} \\
  \hline
  B   & 2011 May 4  & 11204     & ($-$340,$-$332) & 30.0$\deg$        & 
        CRISP   & \Halpha & [0,+1.0]  & ---         & 16.1        & 07:54\,--\,09:54 \\
  {}  & {}          & MF        & \fov{53}{52}    & (0.87)            & 
        AIA & \multicolumn{3}{l}{1600 / 1700} & 24.0        & {} \\
  \hline
  C   & 2015 Jun 19 & 12371     & ($-$539,162)    & 36.6$\deg$        & 
        CRISP   & \Halpha & $\pm$1.5  & 200\,--\,300& 26.7        & 07:15\,--\,08:45 \\ 
  {}  & {}          & EFR        & \fov{60}{59}              & (0.80)            & 
        AIA & \multicolumn{3}{l}{1600 / 1700} & 24.0        & {} \\
  \hline
  D   & 2015 Sep 27 & 12423     & (767,$-$217)    & 56.4$\deg$        & 
        CRISP   & \Halpha & $\pm$1.5  & 200\,--\,300& 32.3        & 07:47\,--\,10:31 \\
  {}  & {}          & EFR        & \fov{70}{57}              & (0.55)            & 
        AIA & \multicolumn{3}{l}{1600 / 1700} & 24.0        & {} \\
  \hline
  E   & 2010 Jun 28 & 11084     & ($-$720,$-$345) & 57.9$\deg$        & 
        CRISP   & \Halpha & $\pm$1.9  & 85          & 22.4        & 08:15\,--\,09:06 \\
  {}  & {}          & MF        & \fov{54}{53}              & (0.53)            & 
        AIA & \multicolumn{3}{l}{1600 / 1700} & 24.0        & {} \\
  \hline
  F   & 2014 Sep 6  & 12152     & (793,$-$268)    & 61.7$\deg$        & 
        CRISP   & \Halpha & $\pm$1.4  & 200         & 11.6        & 08:23\,--\,10:24 \\
  {}  & {}          & MF/DAR        & \fov{54}{55}             & (0.47)            & 
        AIA & \multicolumn{3}{l}{1600 / 1700} & 24.0        & {} \\
  \hline
  G   & 2012 Jun 9  & 11497     & (794,$-$335)    & 65.9$\deg$        & 
        CRISP   & \Halpha & $\pm$2.1  & 86\,--\,258 & 18.4        & 07:29\--\,08:34 \\
  {}  & {}          & MF/DAR        & \fov{55}{55}    & (0.41)            & 
        AIA & \multicolumn{3}{l}{1600 / 1700} & 24.0        & {} \\
  \hline
  H   & 2010 Jun 27 & 11084     & ($-$802,$-$339) & 67.3$\deg$        & 
        CRISP   & \Halpha & $\pm$1.7  & 85          & 17.0        & 13:31\,--\,13:58 \\
  {}  & {}          & MF        & \fov{56}{56}    & (0.39)            & 
        AIA & \multicolumn{3}{l}{1600 / 1700} & 24.0        & {} \\
  \hline
  I   & 2012 Jun 20 & 11504     & (821,$-$319)    & 69.0$\deg$        & 
        CRISP   & \Halpha & $\pm$2.1  & 86\,--\,258 & 18.4        & 07:31\--\,08:17 \\
  {}  & {}          & MF/DAR        & \fov{54}{54}    & (0.36)            & 
        AIA & \multicolumn{3}{l}{1600 / 1700} & 24.0        & {} \\
  \hline
  J  & 2013 Jul 4  & 11778     & (915,$-$126)    & 78.3$\deg$        & 
        CRISP   & \Halpha & $\pm$2.1  & 86\,--\,258 & 8.6         & 10:13\,--\,11:06 \\
  {}  & {}          & EFR        & \fov{57}{57}              & (0.20)            & 
        AIA & \multicolumn{3}{l}{1600 / 1700} & 24.0        & {} \\
  \hline
        \hline
\end{tabular}
\tablefoot{
    The abbreviations below the active region numbers (third column) indicate
    the type of target as discussed in Section~\ref{sec:observations_datasets}:
    EFR (emerging flux region), DAR (decaying active region) and MF (moat flow).
    The eighth column ({\it Range\/}) indicates the extent of the spectral scan with
    CRISP (\ie\ the outermost points with respect to \Halphawav\ line centre).
    
}
\end{center}
\label{tab:datasets}
\end{table*}%

\begin{figure*}[h]
  \centerline{\includegraphics[width=0.97\textwidth]{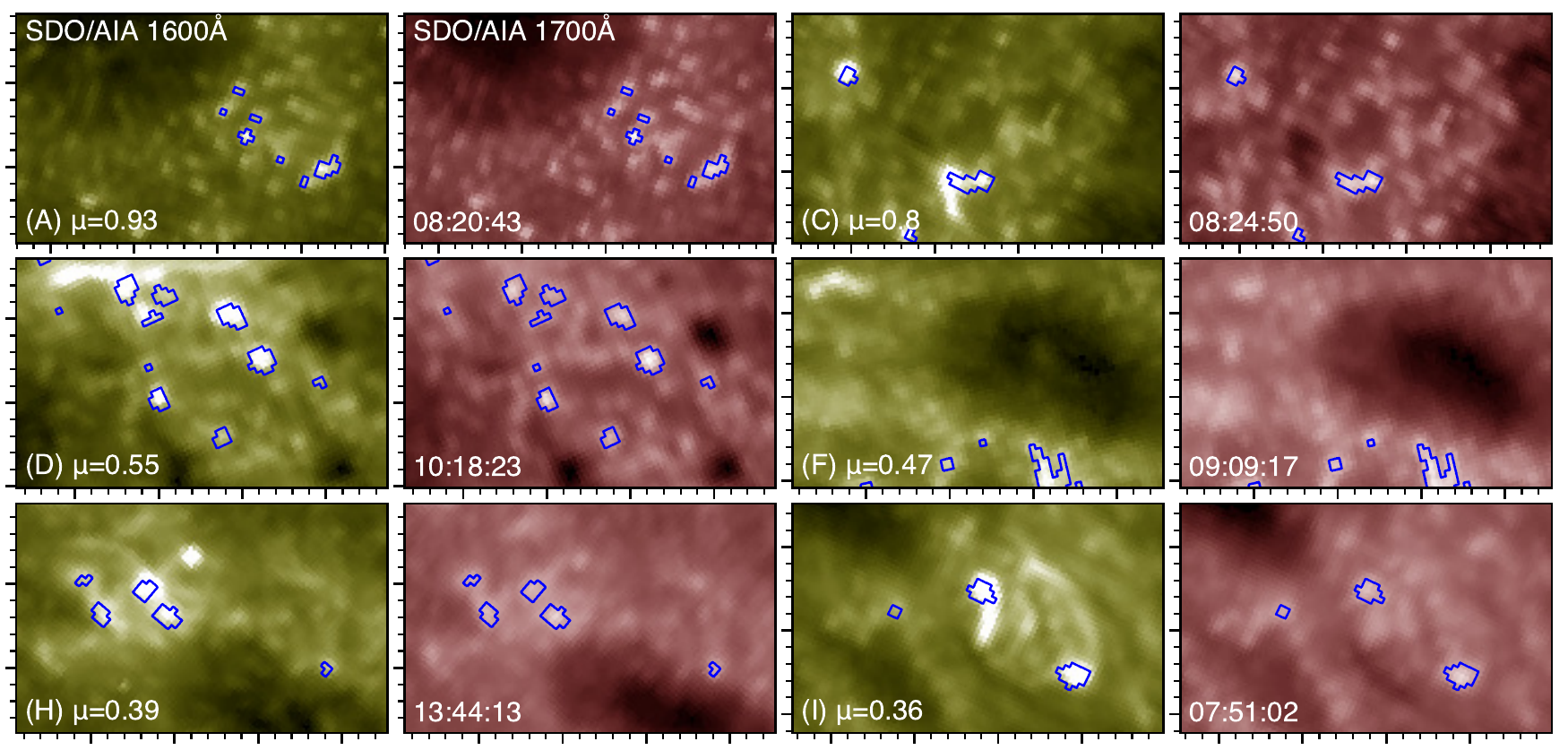}}
  \caption[]{\label{fig:aia_ebfaf} %
  Feature visibilities in AIA 1600\,\AA\ ({\it first and third
  columns\/}) and 1700\,\AA\ ({\it second and fourth columns\/}).
  The panel pairs show selected cutouts for data sets A, C, D, F, H and I,
  labelled with their $\mu$ values in the 1600\,\AA\ panels. 
  The cutout locations are outlined by dashed frames in the
  corresponding panels of Fig.~\ref{fig:fovs}.
  In order to accommodate the dynamic range the square root of the
  intensity is shown. 
  The byte scaling is common between pairs and is defined so that
  quiet areas obtain the same apparent average brightness in 1600 and
  1700\,\AA\ at a scale that saturates at 15$\sigma$ above this quiet
  average for 1700\,\AA\ 
  except for set F where 10$\sigma$ 
  was used to for better display contrast.
  At these values no 1700\,\AA\ image is clipped; only the 
  brightest 1600\,\AA\ features are.
  The blue contours outline 1700\,\AA\ areas at least 5$\sigma$ above
  the quiet average.
  Major tick marks are spaced 10\arcsec\ apart, minor tick marks
  2\arcsec\ as in Fig.~\ref{fig:fovs}.
  }
\end{figure*}

In summary, we aim here to establish the combination of feature
parameter values (brightness, lifetime, area, etc.) applicable to
mid-ultraviolet AIA images that provides optimal recovery of
\Halpha-detected \EBs.
The remainder of this publication is structured as follows. 
The observations and data reduction are described in
Section~\ref{sec:observations}, the analysis method in
Section~\ref{sec:analysis}.
In Section~\ref{sec:results} we present the results followed by
discussion (Section~\ref{sec:discussion}) and conclusions
(Section~\ref{sec:conclusion}). 
The latter end with recommendations for \EB-detection in the AIA
database which represent our ``take-away'' message.

\section{Observations and Data Reduction}\label{sec:observations}
\subsection{SST data acquisition and reduction}
\label{sec:observations_datasets}
In this study we use ten data sets from the CRisp Imaging
Spectropolarimeter (CRISP;
\citeads{2008ApJ...689L..69S}) 
at the SST
for which AIA 1600\,\AA\ and 1700\,\AA\ data are also available.
They are detailed in Table~\ref{tab:datasets}.
Their outstanding quality benefited much from the SST's
adaptive-optics wave-front correction system
\citepads{2003SPIE.4853..370S} 
and from further image reconstruction using Multi-Object Multi-Frame
Blind Deconvolution (MOMFBD;
\citeads{2005SoPh..228..191V}). 
Data sets C, D and F were reduced using the CRISPRED processing
pipeline of \citepads{2015A&A...573A..40D} 
while the remainder was processed using a predecessor of this 
framework.  

The \Halphawav\ line was typically observed with wavelength scans out to
$\pm$1.5--2.1\,\AA, with fixed wavelength spacing for half of the data sets,
while the other half had denser sampling in the core but sparser sampling in the
outer wings (cf.~the eighth ({\it Range\/}) and ninth ($\Delta \lambda$) columns
in Table~\ref{tab:datasets} for further details).
All but the last data sets were complemented with \CaIR\ observations. 
Sets A, C--G and I also included full Stokes polarimetry in \FeI\
6301.4\,\AA\ (sampling only one wing position at $-$0.048\,\AA\ except
for set C where the line was scanned out to $\pm$0.6\,\AA).
However, in this study we use only the \Halpha\ data.

All observations targeted active regions in various stages of their
evolution; their numbers are specified in the third column of
Table~\ref{tab:datasets}. 
In his discovery paper \citetads{1917ApJ....46..298E} 
described his bombs (he called them hydrogen bombs) as exclusively
occurring near sunspots in emerging complex active regions, but they
are also seen near actively flux-shredding sunspots in strongly
decaying active regions with similar serpentine field bundles as in
emerging-flux regions, also producing moving magnetic features
(\citeads{1973SoPh...28...61H}). 
The \EBs\ in data sets A, B and E--I can be generally considered as
moat flow events, with moat flow defined as an organised streaming
motion near a sunspot.
Those in F, G and I were around sunspots in generally decaying active
regions, while those in sets C, D and J occurred in---or as part
of---(recently) emerging flux.

In addition, we note that (1) data sets A, E and H cover (in reverse
order) the same sunspot while it rotated over the disk displaying
stable levels of average activity throughout (no reported flares), (2)
set C was obtained in a highly complex active region with on-going
flux emergence (here observed only two days after it received its NOAA
AR number, a time in which its total sunspot area grew by over 420\%
and it produced one M-class and 9 C-class flares), (3) set D covered
the trailing part of an active region while in the leading part an
M-class flare went off towards the end of our observation, and (4) set
I targeted an active region that, while decaying, was still relatively
complex containing three sunspots.
Finally, we note that set E was previously analysed by
\citetads{2013ApJ...774...32V}, 
set F by \citetads{2015ApJ...811L..33V} 
in studying penumbral microjets, and set J by
\citetads{2016A&A...592A.100R} 
in studying QSEBs far away from the active region.
In addition, SST/CRISP observations of active region NOAA AR 11504 (data set I)
from one day later were analysed in 
\citetads{2015ApJ...798...19N} 
and
\citetads{2015ApJ...805...64R}. 

\subsection{SDO data collection and co-alignment}
Corresponding SDO image cutout sequences for the ten SST/CRISP data
sets were downloaded, precisely cross-aligned (all AIA channels to
HMI), and co-aligned with the SST images using an IDL pipeline
developed by the third author.
It is available at his
website\footnote{\url{http://www.staff.science.uu.nl/~rutte101/rridl/00-README/sdo-manual.html}}
and will be detailed elsewhere.
For each SST data set its product consists of eleven HMI and AIA image
cutout sequences that are rotated to the SST image orientation and
resampled to be precisely co-spatial (to within 0\farcs{1}) and as close as possible in time
(through nearest-neighbour frame selection) with the SST
images. 
The SDO data were interpolated to the ten times finer SST pixel scale
using nearest-neighbour sampling to maintain the original AIA pixel
shapes for determining area constraints on their native scale.
The AIA EUV sequences are not used in this analysis, only the
mid-ultraviolet (1600\,\AA\ and 1700\,\AA) and HMI ones.

Figure~\ref{fig:fovs} shows co-aligned sample images in CRISP \Halpha\
and AIA 1700\,\AA\ from all ten data sets. 
Generally, there is good spatial correspondence between the brightest
features in each pair, but there are also many differences.
Figure~\ref{fig:aia_ebfaf} shows sample comparisons of AIA 1600 and
1700\,\AA\ image cutouts for four data sets at the same sample times
as in Fig.~\ref{fig:fovs} to illustrate differences between these AIA
diagnostics. 
In this figure each panel is not byte-scaled individually with its own
saturation clip as done for best scene visibility in
Fig.~\ref{fig:fovs}, but each pair shows the square root of the
intensity at a common range set by requiring that the quiet-area
averages defined by the masks defined below (taken over the whole time
sequence) obtain the same apparent brightness, with the same
high-level saturation cutoff per pair and without clipping the
1700\,\AA\ images.

The quiet parts appear very similar between 1600 and 1700\,\AA,
showing nearly identical bright-grain patterns. 
The internetwork hearts between these appear darker in 1600\,\AA\
which is probably due to longer exposure (about 3\,sec instead of
1\,s) that causes more smearing of the rapidly moving filamentary
weak-brightness patterns set by interfering acoustic shocks.

The scaling also makes clear that both \EBs\ (bright and roundish in both)
and FAFs (elongated bright features in 1600\,\AA\ not present in
1700\,\AA) reach higher contrast over the quiescent network in
1600\,\AA, presumably from \CIV\ contributions. 
They also appear slightly larger, presumably from scattering.
The common 5$\sigma$ above-quiet-average 1700\,\AA\ brightness
contours in Fig.~\ref{fig:aia_ebfaf} illustrate one ingredient of the
\EB\ detection recipe developed below.

\begin{table*}[h]
\caption{\Halpha\ intensity, area and lifetime thresholds for \EB\ selection and resulting detection rates in
recent literature.}
\begin{center}
\begin{tabular}{llllllll}%
        \hline \hline
  {}                            & \multicolumn{4}{c}{Intensity threshold} & Size
  & Lifetime & {} \\
  \cline{2-5}
  {}                            & Single/ & Contrast value(s) & \Halpha\ wing & Reference &
  threshold & threshold & Detection rate \\
  Study                         & Double  & ($I_{\rm{EB}}/I_{\rm{ref}}$) &
  [\AA]&{} &
  [\sqasec] & [sec] & [arcmin$^{-2}$ min$^{-1}$] \\
        \hline
  \citeads{1987SoPh..108..227Z}\tablefootmark{a} & Single  & 1.28       &
        $-$0.75 / $-$1.0 & Sub-FOV & 0.6 & 480 & 12.50 \\
  \citeads{2002ApJ...575..506G} & Single  & 1.05 / 1.08 / 1.20          & 
        $-$0.8 & Per pixel & N/A & N/A & 5.76 / 2.86 / 0.28 \\
  \citeads{2011ApJ...736...71W}\tablefootmark{b,c} & Single  & 1.16 / 1.27& 
        $\pm$(0.9--1.1) & Sub-FOV &  0.025 & 240 & 1.42 \\
  \citeads{2013SoPh..283..307N} & Single  & 1.30                        & 
        $\pm$0.7 & Full FOV &  0.037 & N/A & 15.49 \\
  \citeads{2013ApJ...774...32V}\tablefootmark{c} & Double  & 1.55 \& 1.40                & 
        $\pm$(0.9--1.1) & Full FOV  & 0.018 & 45, 55 & 1.92, 1.30 \\
  \citeads{2015ApJ...798...19N} & Single  & 1.50                        & 
        $\pm$(0.9--1.2) & Nearby QS & 0.014 & N/A & 0.79 \\
        \multirow{2}{*}{\citeads{2015ApJ...812...11V}\tablefootmark{c,d}} &
        \multirow{2}{*}{Double} & 1.55 \& 1.40 & 
        \multirow{2}{*}{$\pm$(0.9--1.1)} & \multirow{2}{*}{Full FOV}  &
        \multirow{2}{*}{0.018} & 22 & 0.82, 1.08 \\ 
        {}&{}& 1.45 \& 1.30 & {} & {}  & {} & 23 &  1.14\\ 
  \citeads{2016ApJ...823..110R}\tablefootmark{e}& Double  & 1.45 \& 1.30                & 
        $\pm$1.0 & Nearby QS & 0.052 & 45 & 1.34 \\
  \citeads{2017GSL.....4...30C}\tablefootmark{f} & Single  & 1.52 & 
        +1.0 & Full FOV  & 0.114 & 100 & 0.55 \\
        Present study\tablefootmark{e}& Double & 1.45 \& 1.30                & 
        $\pm$(0.9-1.1) & Masked QS & 0.035 & 60 & 1.11 \\
        \hline
\end{tabular}
\tablefoot{
  The fourth column ({\it \Halpha\ wing\/}) specifies the wavelength offset with
  respect to line centre used for \EB\ identification.
  Values in the last two columns that correspond to different data sets within a
  study are comma-separated, while in the last column the rates for different
  contrast thresholds are separated by slashes (/). 
  \newline
  \tablefoottext{a}{The values in this study are not formal thresholds, but rather the average of manually selected
  events.} 
  \tablefoottext{b}{This study specified a threshold of 3$\sigma$ (for
  non-plage) and 5$\sigma$ (for plage) above the local average of the considered
  sub-fields-of-view, corresponding to the contrast values given
  here.}
  \tablefoottext{c}{Intensity thresholding was performed on the
    combined wing-average of the blue and red wings ($I_w$ as defined in
  Section~\ref{sec:hawings}).}
  \tablefoottext{d}{This study specified different intensity thresholds for two data sets
  versus the third, here split into two rows where values differ.}
  \tablefoottext{e}{Intensity thresholding was performed on the blue and red
  wings separately.}
  \tablefoottext{f}{This study specified a threshold of 4$\sigma$ above the average, which
  corresponds to the contrast value given here (Yajie Chen, private communication).}
}
\end{center}
\label{tab:thresholds_literature}
\end{table*}

\section{Analysis methods} \label{sec:analysis}
\subsection{Automated detection with \EBDETECT} 
\label{sec:analysis_detection}
Our aim is to establish the optimal recipe to retrieve
\Halpha-detected \EBs\ from concurrent AIA ultraviolet images.
We therefore first detect \EBs\ in the CRISP \Halpha\ data and then
use these to evaluate the recovery success of various AIA detection
criteria including finding which AIA passband works best.

Our detection code \EBDETECT\ (written in IDL) used for both the SST
and the AIA data builds on four key elements:
\begin{itemize}  \itemsep=1ex

\item {\it Brightness thresholds\/}. 
Initial identification is done by selecting pixels passing a specified
intensity threshold. 
For \Halpha\ a double intensity-threshold criterion (on wing-average images as
defined further down) serves to
recognise both the high-intensity fine-structure kernel and the
surrounding lower intensity halo.
These thresholds are expressed in the average intensity and the
standard deviation around that for all pixels in quiet areas of the
field of view over the full sequence duration as defined below.

\item {\it Size constraints\/}. 
A minimum area is set to prevent selecting single-pixel features that
are likely spurious signals, while setting a maximum prevents
picking up extended regions of plage (a particular issue for the AIA
images at low brightness threshold).
  
\item {\it Continuity constraints\/}. 
Detections are subsequently checked for overlap between sequential
frames, requiring at least one pixel area (native size, \ie\
$0.6\arcsec\times0.6\arcsec$ for AIA, $0.059\arcsec\times0.059\arcsec$
for SST) overlap from frame to frame.
However, in order to alleviate bad seeing moments there may be
intermediate gaps of durations up to the minimum-lifetime constraint
(\ie\ up to $\sim$60\,s).

\item {\it Lifetime constraint\/}. Finally, those detections that have
passed the above hurdles must also live longer than 1\,min.

\end{itemize}

\EBDETECT\ builds on criteria established in
\citetads{2011ApJ...736...71W}. 
Earlier versions were used in
\citetads{2013ApJ...774...32V}, 
also for detection in \CaIR, and
\citetads{2015ApJ...812...11V} 
while \citet{2018Vissersuvbinv} 
employ the current version.  
The key changes from our earlier versions are (1) the values of the
brightness thresholds and how these are determined, and (2) for
\Halpha\ thresholds are now applied to the blue and red wings
separately.
We first detail these changes.

\paragraph{Reference intensities for brightness thresholding.} 
\label{sec:quietprof} The first modification serves to define
dataset-independent brightness thresholds.
In our previous \Halpha\ studies a double brightness threshold of
155\% and 140\% of the average brightness generally yielded good
results, but this average was evaluated over the full fields of view.
This had to be amended for cases where the umbral and/or penumbral
areas were relatively small by varying the two thresholds over
160--145\% and 145--130\%, respectively.
Such target- and reference-dependent variations are not uncommon, as
shown by Table~\ref{tab:thresholds_literature} which lists thresholds
used during the past 40 years, but they hinder the definition of a
general recipe. 
We therefore now define thresholds no longer with respect to the
average over the full field of view, but only over its quiet areas,
\ie\ excluding sunspots, pores and significantly bright plage.
In several studies the \Halpha\ wing enhancement was normalised by
nearby quiet-area averages, but here we average over all quiet pixels
in each field of view to obtain better statistics. 
This approach requires automated definition of blocking masks.

\begin{figure*}[bht]
  \centerline{\includegraphics[width=\textwidth]{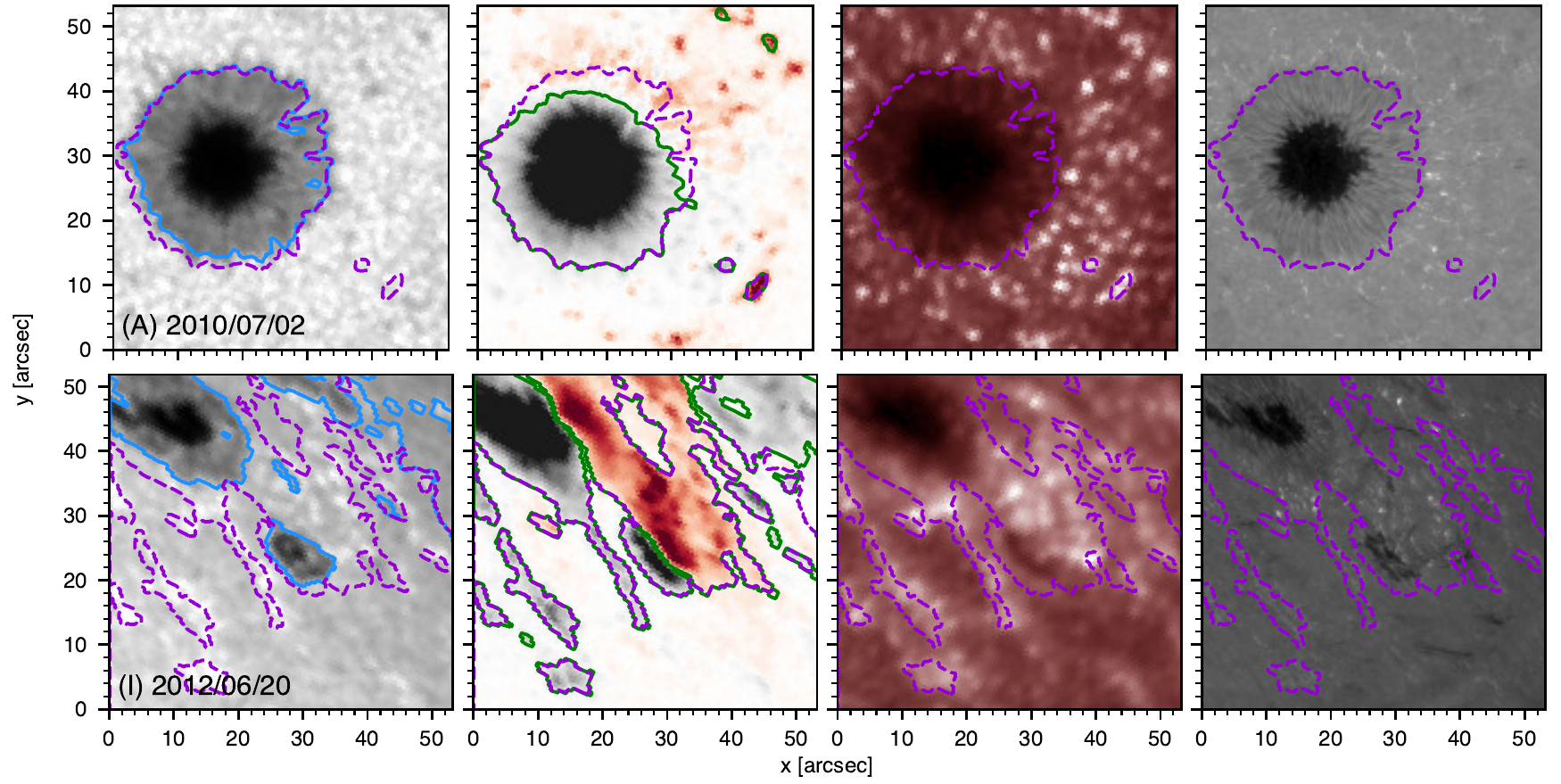}}
  \caption[]{\label{fig:hmimasking} %
  HMI mask construction for data sets A (viewing angle $\mu=0.93$, {\it
  top row\/}) and I ($\mu=0.36$, {\it bottom row\/}).
  The masks serve to define the local quiet area for reference brightness
  thresholding. 
  {\it Left to right\/}: HMI continuum image, HMI line-of-sight
  magnetogram (positive/negative polarity in red/black, zero field
  strength white), AIA 1700\,\AA\ image, blue wing \Halpha\ image.
  The HMI intensity mask is outlined by blue contours in the first
  column, the HMI magnetic field mask by green contours in the second
  column. 
  The composite mask is shown in all panels by purple dashed contours.
  The small green-only islands in the second column are blocked by the
  minimum-area constraint. 
  In the last panel the blocked part covers most of the upper half
  including all \EBs.
  }
\end{figure*}

A straightforward approach would seem to mask out the stronger-field
magnetic areas on co-aligned HMI line-of-sight magnetograms, but
substantial offsets between HMI magnetogram contours and HMI
continuum-image contours can occur away from disc centre.
We therefore define a composite mask by first thresholding the HMI
continuum images at 60\% of their maximum intensity to discard darker
sunspots and pores, and then combine this low-intensity block with
magnetogram blocking above $|B_{\rm los}| = \mu\,\times\,180$~Gauss
which also removes bright plage.  
Such masks are determined for every image and then multiplicatively
compounded into a single composite mask used for the full sequence so
that a pixel blocked at any time gets blocked at all times.
Any passed feature smaller than 4 AIA pixel-equivalent area (about 400
SST pixels) is then also blocked, as are those smaller than 60 AIA
pixel-equivalent area if they lie isolated within a blocked region
(\eg\ a bright penumbral feature).
Figure~\ref{fig:hmimasking} shows results from this procedure. 

For data set J, which contains the limb in the field of view, we
applied an additional limb mask blocking the off-limb pixels and also
the outer $\sim$5\arcsec\ of the disc because its large radial
intensity drop strongly influences the mean value.

\paragraph{\Halpha\ wing treatment.}\label{sec:hawings}
The second modification addresses extreme Doppler shifts of \Halpha\
that are imposed by fast flows in overlying canopy fibrils.
In previous studies we used wing-average images constructed as $I_w \equiv
(I_b + I_r) / 2$, where the blue-wing $I_b$ and red-wing $I_r$ are the spectral
averages over three wavelength tuning positions centred at $-$1\,\AA\ and
+1\,\AA, respectively (\ie\ effectively $\pm$(0.9--1.1)\,\AA).
However, in the presence of strong canopy Doppler shifts using such mean value
combining both wings can put the intensity below the threshold and so reduce the
apparent area of \EB\ candidates or ignore them altogether. 
To account for these effects we therefore apply the brightness thresholding to
the wing-averages $I_b$ and $I_r$ separately, \ie\ a pixel need only pass the
threshold in one of the wings to still carry over to the next step 
(an approach similar to the one in 
\citetads{2016ApJ...823..110R}, 
except that they also included a line core constraint).
In the remainder we refer to $I_b$ and $I_r$ as wing-average images.
Note that due to larger spacing of the wavelength sampling in datasets C, D and
F, for those cases this averaging effectively spans $\pm$(0.8--1.2)\,\AA.

\subsection{Parameter values for \EB\ detection in \Halpha}
\label{sec:ebdetect_halpha_params}
For our \EB\ detections in \Halpha\ wing-average images we ended up with the
following constraints: (1) a double brightness threshold of 145\%
(core) and 130\% (halo) over the quiet-Sun average which must be
exceeded in at least one of the wings, where halo pixels are adjacent
to already defined core or halo pixels, (2) a minimum area of 10
connected core-plus-halo SST pixels (corresponding to a linear extent
of 0\farcs{2}--0\farcs{6} depending on feature elongation, about
0.035\,\sqasec), and (3) a minimum lifetime of about 60\,s but
allowing non-detection gaps up to about 60\,s to accommodate
bad-seeing instances. 
The latter time constraints translate into 2--7 frames depending on
the observing cadence and effectively span 53--68\,s. 

Two further adjustments were made to account for particular data-set
peculiarities. 
Firstly, in set A we found that the average intensities varied strongly
in time, by
nearly 12\% between the first and last frames as compared to 0--4\%
for the other data sets.
We compensated for this variation by taking a running mean with a
boxcar of about 5\,min
(equivalent to 13 frames, spanning times over which the mean
intensities changed less than 1\%).
This correction resulted in detecting events that were missed
previously, especially in the beginning of the time sequence.

Secondly, the wing-average images of data sets I and J, which are the
most limbward ones, were more strongly affected in the \Halpha\ wings
at $\pm$(0.9--1.1)\,\AA\ by Dopplershifts of overlying canopy fibrils,
which we remedied by moving the sampling wavelengths for the average
taking outward to $\pm$(1.0--1.2)\,\AA.
This correction resulted in fewer dubious small-scale weak detections.

Our \Halpha\ constraints resulted in the detection of 1735 candidates
in total from our ten SST data sets. 
We verified their nature by visual inspection of the resulting \EB\
contours overlaid on \Halpha\ wing-average images using CRISPEX and
concluded that, even though there is a comparatively large population
of very small-scale, short-lived events in data sets C, D, F and J for
which identification is less obvious, at least 90\% of the 1735
automated \Halpha\ detections represented bona fide \EBs. 
We also found that our recipe recovers over 94\% of what we recognise
as bona fide \EBs.

\subsection{Parameter grid for \EB\ detection with AIA}
We applied \EBDETECT\ to the AIA 1600 and 1700\,\AA\ sequences for a
grid of parameter values that were varied independently.
Firstly, eight brightness thresholds were considered, varying at
1$\sigma$ steps between 3$\sigma$ and 10$\sigma$ above the average
quiet-Sun intensity and its standard deviation $\sigma$ determined
using the same HMI-based masks as for \Halpha. 
Initial tests with thresholds at only 1$\sigma$ and 2$\sigma$
mis-identified too much normal network for any combination of the
other parameters (as expected from the \Halpha-1700\,\AA\ scatter
plots in Fig.~7 of \citeads{2013ApJ...774...32V}), 
so we restricted the range to values from 3$\sigma$ upwards.  

Secondly, nine different area constraints were applied: a minimum of 1 AIA
pixel-equivalent with a 3, 6, 9, \ldots, 24 and 27 AIA-pixel maximum.
Since the AIA data were rescaled to the CRISP pixel size (while
retaining the AIA pixel shapes, \ie\ without interpolation) one AIA
pixel corresponds to roughly 100-110 CRISP pixels; to be sure to catch
single AIA pixels we lowered this value to 95 CRISP pixels as AIA
pixel-equivalent for the lower limit while assuming 110 CRISP pixels
as AIA single-pixel-equivalent for the upper limit.

Lastly, six different lifetime constraints were set: (1) a 1\,minute minimum and
no maximum, and (2) 1\,minute minimum and a maximum of 5, 10, 15, 20 and
30\,min, respectively.
The continuity criteria for the \Halpha\ detections were maintained:
at least one native pixel overlap between frames while permitting up
to about 1\,min of non-detection (the latter only for consistency with
the \Halpha\ formalism since AIA does not suffer seeing variations).

\begin{figure*}[bht]
  \centerline{\includegraphics[width=\textwidth]{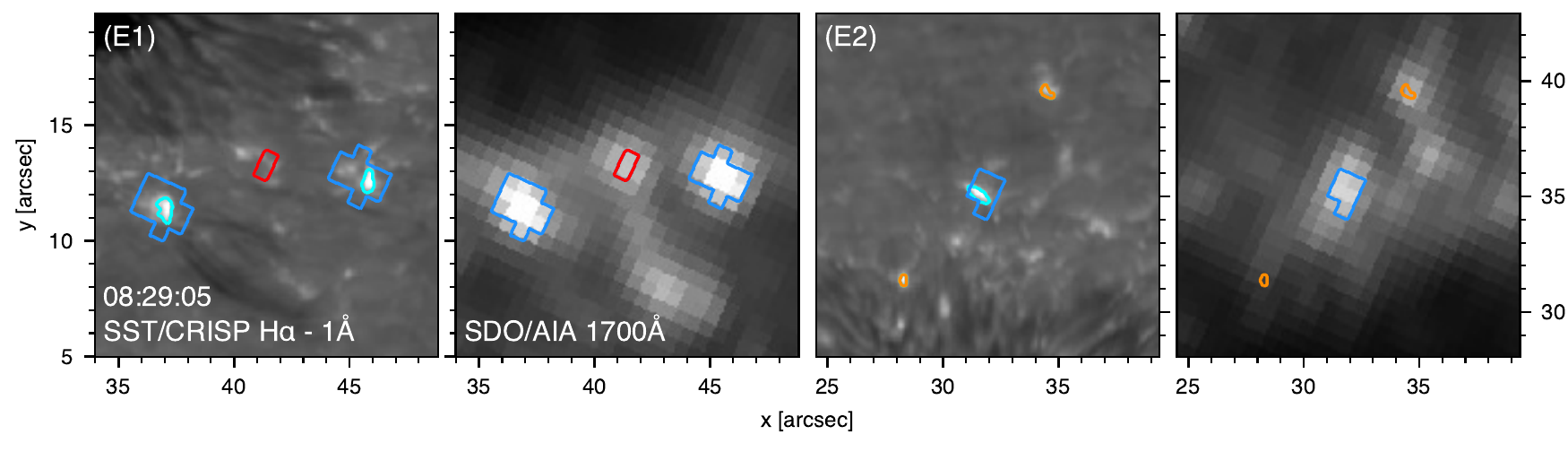}}
  \caption[]{\label{fig:detection_comparison} %
  Examples of detection evaluations from data set E.
  The two panel pairs show \Halpha\ $-$(0.9--1.1)\,\AA blue-wing
  average intensity and AIA
  1700\,\AA\ intensity for the similarly numbered cutouts in panel
  pair E of Fig.~\ref{fig:fovs}.
  Detection contours in \Halpha\ ({\it cyan and orange\/}) and AIA 1700\,\AA\
  ({\it blue and red\/}; 5$\sigma$ threshold, minimum of 1 AIA px and 2 frames
  visibility) are overlaid. 
  The blue AIA contours (all panels) are classified as true positive (TP)
  detections from their degree of overlap with the cyan \Halpha\
  contours (first and third panel), whereas the red contour in the first panel pair shows a
  false positive (FP), the orange contours in the second panel pair two
  false negatives (FN).
  }
\end{figure*}

\subsection{Correspondence evaluation using performance metrics} 
\label{sec:analysis_completeness}
The next step is to compare the results of the AIA detection grid to
the \EB\ detections in \Halpha\ where we assume the latter to be all
correct and also complete, \ie\ that no \EBs\ went undetected.
Of course, any automated detection code searching specific features
must miss some and misidentify others, but we discard such errors for
\Halpha\ on the basis of our visual checks.

For each AIA detection we then established whether there is overlap in
time and space with any or multiple \Halpha\ detections, requiring
overlaps during at least half the lifetime and half the area of one of
the two detections.
Tests where these requirements were varied down to only one third
overlap and up to three quarters overlap in area and lifetime
suggested no significant differences.
Thus, for each AIA detection we so found whether it was correct (and
if so, with how many \Halpha\ detections it overlapped) and added to
the {\it true positive\/} (TP) AIA score or instead to the erroneous
{\it false positive\/} (FP, without \Halpha\ counterpart) AIA score.
The {\it false negative\/} (FN) score then remains as counting SST
\Halpha\ detections without AIA counterpart.
Table~\ref{tab:contingency} visualises these in a contingency matrix,
while Fig.~\ref{fig:detection_comparison} shows examples of valid TP
detections and FP and FN error cases.

\begin{table}[h]
  \caption{Contingencies for detection correspondence in
  \Halpha\ and AIA.}
\begin{center}
\begin{tabular}{l|ll}%
        \hline \hline
        Event detected & \multicolumn{2}{c}{Event detected in \Halpha?} \\
        \cline{2-3}
        in AIA?& Yes & No \\ \hline
  Yes & true positive (TP)  & false positive (FP) \\ 
  No  & false negative (FN)  & N/A \\
  \hline
\end{tabular}
\tablefoot{No detection in both \Halpha\ and AIA (\ie\ lower right) is
technically a true negative (TN), but cannot be quantified for our case.}
\end{center}
\label{tab:contingency}
\end{table}

To measure the success of a particular parameter combination we use
the precision P defined as the fraction of \Halpha\ detections in all
AIA detections
\begin{equation}\label{eq:precision}
  \rm{P} = \frac{\rm{TP}}{\rm{TP+FP}},
\end{equation}
and the recall R defined as AIA's recovery fraction of all \Halpha\
detections
\begin{equation}\label{eq:recall}
  \rm{R} = \frac{\rm{TP}}{\rm{TP+FN}}.
\end{equation}
A metric combining these is their equally-weighted harmonic mean:
\begin{equation}\label{eq:f1}
  \rm{F}_{1} =  \frac{2 \times \rm{P} \times \rm{R}}{\rm{P + R}} = 
  \frac{2 \times \rm{TP}}{2 \times \rm{TP + FP + FN}}
\end{equation}
which peaks where P and R are both high.

A crucial decision is what type of optimisation is desired. 
If one wishes to recover as many \EBs\ in AIA data as possible one
should maximise the recall fraction R at the expense of the precision
P, but if one instead desires that as many as possible of the
AIA-detected events are \Halpha-verified \EBs\ the precision P should
be maximised at the expense of the recall R.
The \fone\ score covers the middle ground by maximising TP while
minimising FP and FN.
The priority choice between these three should be defined by the
nature of the particular application.
Here we present all three but focus on P and \fone\ because optimising
R is not realistic at the tenfold SST--AIA resolution difference.

\section{Results} \label{sec:results}
\subsection{\Halpha\ results} \label{sec:detections_halpha}
\begin{figure*}[bht]
  \centerline{\includegraphics[width=\textwidth]{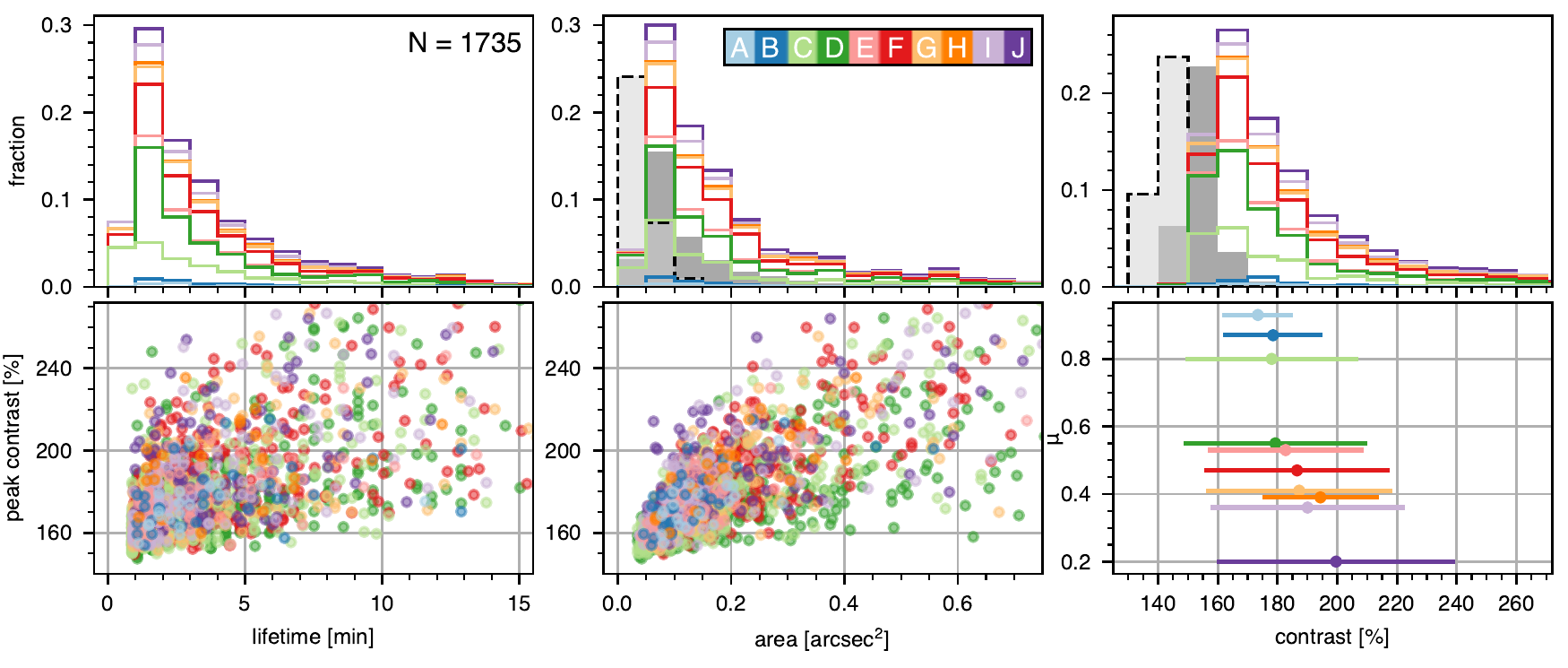}}
  \caption[]{\label{fig:halpha_stats} %
  Statistics of all 1735 \EB\ detections in \Halpha.
  In all panels the ten different data sets are colour-coded
  blue-green-red-orange-purple along increasing viewing angle
  (decreasing $\mu$); see also the top middle panel for the
  colour-correspondence of each data set A--J.
  {\it Upper row:\/} stacked occurrence histograms as function of
  lifetime, maximum area, and contrast.
  The histogram bin sizes are 1\,min, 0.05\,\sqasec, and 10\%.
  Since the coloured histograms are stacked the topmost dark purple ones
  also outline the cumulative distributions. 
  The dark grey and dash-outlined, light grey overlays in the
  second and third panels show the cumulative distributions for the
  mean and minimum area and contrast over the lifetime of each
  detection, respectively.
  Note that these grey overlay histograms have been scaled down by a factor
  3 in height to fit on the same scale as the coloured histograms.
  {\it Lower row:\/} scatter plots of peak contrast as function of
  lifetime and maximum area.
  The last panel shows peak contrasts at the different $\mu$ values,
  with the bar lengths showing the rms peak contrast spreads around
  their mean values shown by the dots.
  }
\end{figure*}

\begin{figure}[h!] 
  \centerline{\includegraphics[width=\columnwidth]{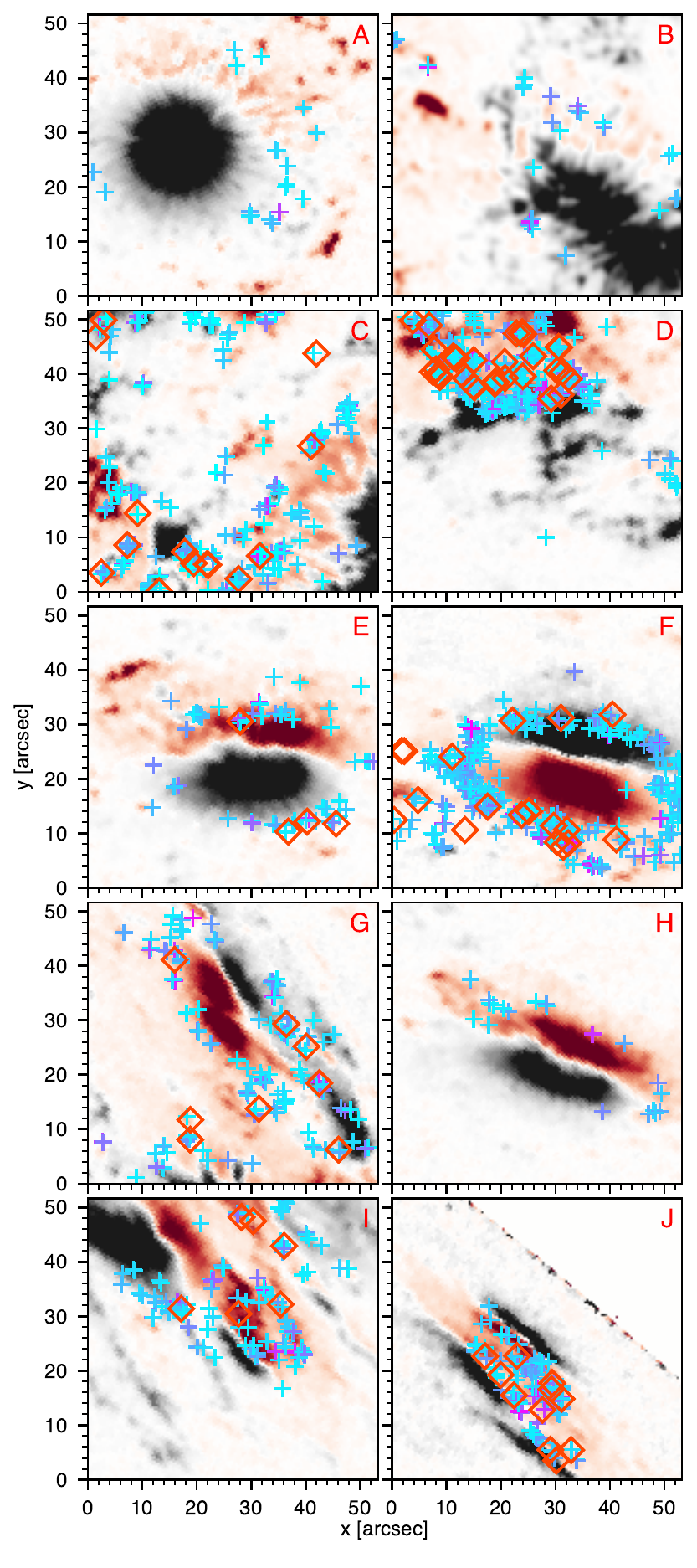}}          
  \caption[]{\label{fig:halpha_locs} %
  Spatial distribution of all \Halpha\ \EB\ detections ({\it coloured
  crosses\/}) for each data set (labelled in the top right of each
  panel) overlaid on HMI line-of-sight magnetograms, with positive
  (negative) polarity in red (black) and zero field strength in white.
  The cross colour is indicative of the event lifetime, ranging from
  cyan to purple for lifetimes between 0--15\,min, with orange open diamonds for
  longer-lived events.                               
  Note that the field-of-view in panel D is slightly shifted with respect to
  Fig.~\ref{fig:fovs} to show all \Halpha\ detections and that the off-limb part
  of panel J has been manually set to zero.
  The magnetogram scales were byte-scaled independently. 
  }
\end{figure}

The \Halpha\ results are summarised in Table~\ref{tab:halpha_stats}
and in Figs.~\ref{fig:halpha_stats} and \ref{fig:halpha_locs}.  
The detection rates (last column of Table~\ref{tab:halpha_stats}) vary
between 0.34--3.00 detections\,arcmin$^{-1}$\,min$^{-1}$, similar to
most studies over the past decade (cf.\ last column of
Table~\ref{tab:thresholds_literature}).
There is no obvious trend with viewing angle; all fall within
1$\sigma$ spread from the average except for set C.

The 1735 detected \Halpha\ \EBs\ have average lifetimes about 3\,min, with the
lifetime distribution peaking closer to half of that but with a
considerable tail out to about 15\,min (third column of
Table~\ref{tab:halpha_stats} and first panel of
Fig.~\ref{fig:halpha_stats}).
About 6\% of the \Halpha\ detections have longer lifetimes, unevenly
spread up to over an hour, but 89\% of the events have a lifetime of
10\,min or less.
Note that these are total lifetimes within the detection constraints.
They include re-brightenings and should not be taken to describe
elemental \EB\ features (\ie\ substructure) which are known to vary on
timescales of seconds or less (cf.~Fig.~3 and the accompanying movie
in \citeads{2011ApJ...736...71W}). 

The average maximum area of the \Halpha-detected \EBs\ lies around
0.15\,\sqasec.  
However, the area distributions (solid coloured outlined histograms in the second panel
of Fig.~\ref{fig:halpha_stats}) peak at small areas
(0.05--0.1\,\sqasec); the mean of the area minima per detection
(dash-outlined, filled light grey overlay) peaks below 0.05\,\sqasec.
\EBs\ are truly sub-arcsecond features requiring the best telescope
resolution presently available.

\begin{table}[h]
\caption{\Halpha\ detection statistics.}
\begin{center}
\begin{tabular}{crccc}%
        \hline \hline
  {}   & \multicolumn{4}{c}{Detections statistics} \\
  \cline{2-5}
  {}  & Number  & Lifetime & Max. Size & Rate \\
  Set & {}      & [min]    & [\sqasec] &[arcmin$^{-2}$ min$^{-1}$] \\
  \hline
  A & 19  & 2.61$\pm$2.25 & 0.13$\pm$0.06 & 0.34 \\ 
  B & 31  & 3.49$\pm$3.13 & 0.11$\pm$0.11 & 0.68 \\ 
  C & 333 & 2.23$\pm$5.83 & 0.11$\pm$0.25 & 3.00 \\ 
  D & 470 & 2.68$\pm$8.22 & 0.14$\pm$0.57 & 1.21 \\ 
  E & 62  & 2.61$\pm$6.13 & 0.13$\pm$0.21 & 0.82 \\ 
  F & 395 & 3.09$\pm$6.30 & 0.16$\pm$0.25 & 1.85 \\ 
  G & 145 & 3.07$\pm$5.47 & 0.16$\pm$0.30 & 1.09 \\ 
  H & 21  & 3.40$\pm$2.87 & 0.20$\pm$0.11 & 0.36 \\ 
  I & 126 & 2.76$\pm$5.57 & 0.13$\pm$0.24 & 1.24 \\ 
  J & 133 & 3.44$\pm$8.24 & 0.16$\pm$0.36 & 0.55 \\ 
  \hline
  avg&---& 2.94$\pm$1.82 & 0.14$\pm$0.09 & 1.11$\pm$0.77 \\ 
        \hline
\end{tabular}
\tablefoot{The third and fourth columns list the median lifetimes and median
maximum areas with their standard deviations.
The last column gives the occurrence rate. 
The last row lists the averages except for the number count (which
would not be meaningful).}
\end{center}
\label{tab:halpha_stats}
\end{table}

The peak contrasts (solid coloured outlined histograms in the third panel of
Fig.~\ref{fig:halpha_stats}) are measured as the maximum \Halpha\
intensity (the brightest pixel in all time steps showing the event)
expressed as percentage brightening over the sequence-averaged mean
intensity of the non-masked quiet parts of the field of view
(Fig.~\ref{fig:hmimasking}). 
The average is close to 180\%. 
The summed distribution shown by the purple histogram peaks in the
160--170\% bin, close to the 169\% average value over the mean
contrast (not its peak but its mean over the detection lifetime)
distribution shown by the filled dark grey overlay.

In the lifetime histograms in the first panel of
Fig.~\ref{fig:halpha_stats} the individual data sets display rather
similar behaviour, independent of viewing angle.
Comparison with the second and third panels shows that where many
short-lived events are detected (\eg\ data sets C (light green), D
(dark green) and F (red)), these are typically also on the small side
and typically weaker in peak contrast.
The scatter diagrams in the first two lower panels of
Fig.~\ref{fig:halpha_stats} confirm the positive correlations of peak
contrast with lifetime and area, with the latter somewhat tighter.

The last panel of Fig.~\ref{fig:halpha_stats} shows the variation of
the mean values and the spread of the peak contrasts per data set
ordered for viewing angle along the vertical $\mu$ axis. 
There is a slight trend to larger peak contrast towards the limb, with
also larger spread. 
The mean peak contrasts reach up to 200\% above the quiet-area
average, much higher than typical mean contrasts (peak of the dark grey
overlay in the upper panel).

Figure~\ref{fig:halpha_locs} shows the spatial distribution of all
\Halpha\ detections overlaid on HMI line-of-sight magnetograms. 
The \EB\ candidates in data sets A, B, E--I are mainly located in
sunspot moat flows. 
In sets D and J they are concentrated between large assemblies of
opposite-polarity fields.
The longer-lived events (total lifetimes above 15\,min, orange
crosses) are predominantly found in areas showing more active-region
complexity and/or flux emergence (\eg\ sets C, D and J), or intense
active-region decay (\eg\ sets F and G).

\subsection{AIA results} \label{sec:completeness}
\begin{figure*}[h]
  \centerline{\includegraphics[width=\textwidth]{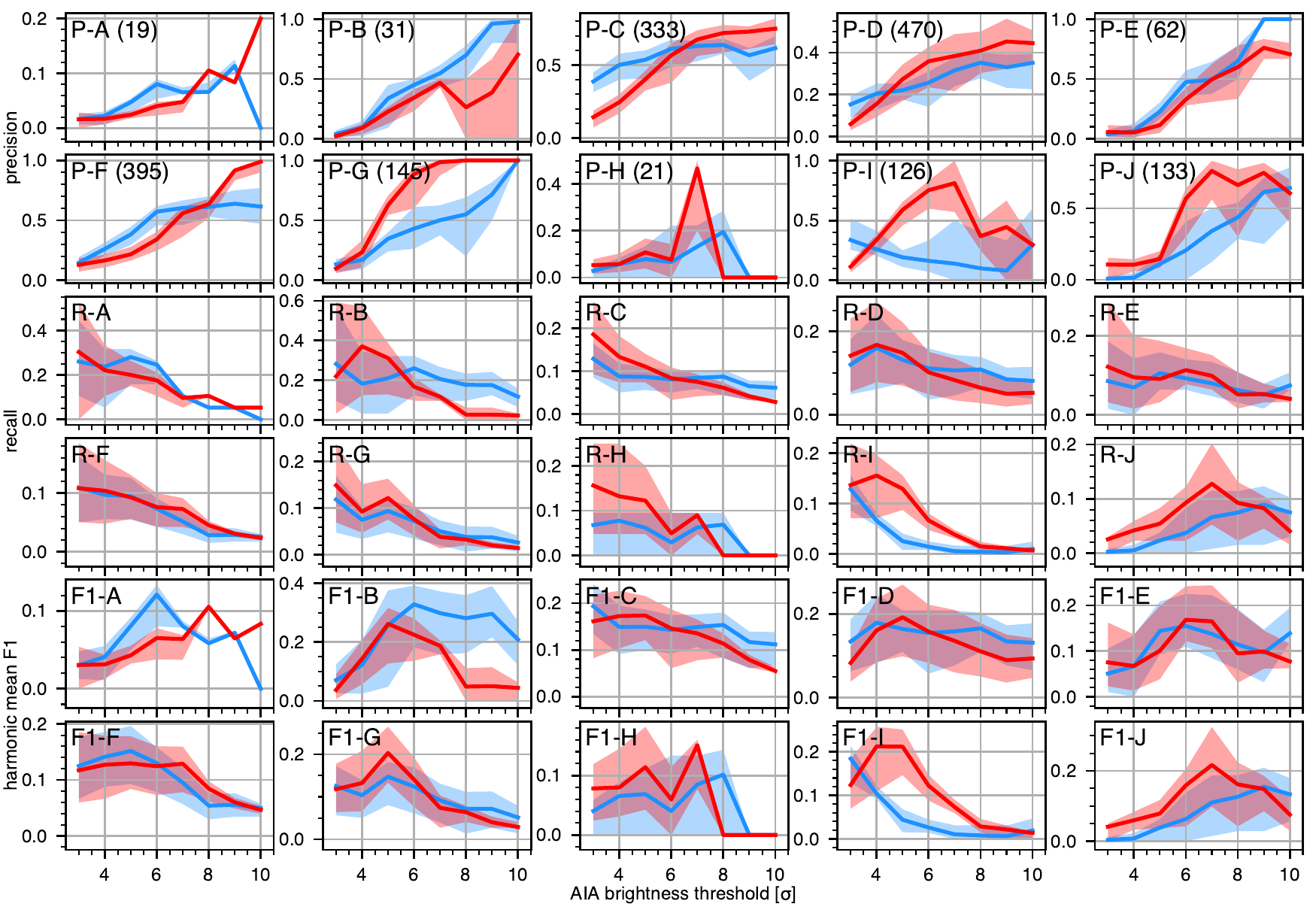}}
  \caption[]{\label{fig:prf_plots_all} %
    AIA performance metrics precision P ({\it top two rows\/}, P-A--P-J),
    recall R ({\it middle two rows\/}, R-A--R-J) and the \fone\ score
    ({\it bottom two rows\/}, F1-A--F1-J) as function of the imposed AIA
  brightness threshold for data sets A--J (specified at top left per
  panel, in the top two rows with the number of \Halpha\ detections).
  In each panel the solid curves trace the averages for 1600\,\AA\
  ({\it blue\/}) and 1700\,\AA\ ({\it red\/}), with correspondingly
  coloured shading (and darker overlaps) showing the spread per
  threshold that results from the different additional area and
  lifetime constraints in the AIA detection grid. 
  The $y$-axis scales differ per panel.
  }
\end{figure*}

\begin{figure*}[bht]
  \centerline{\includegraphics[width=\textwidth]{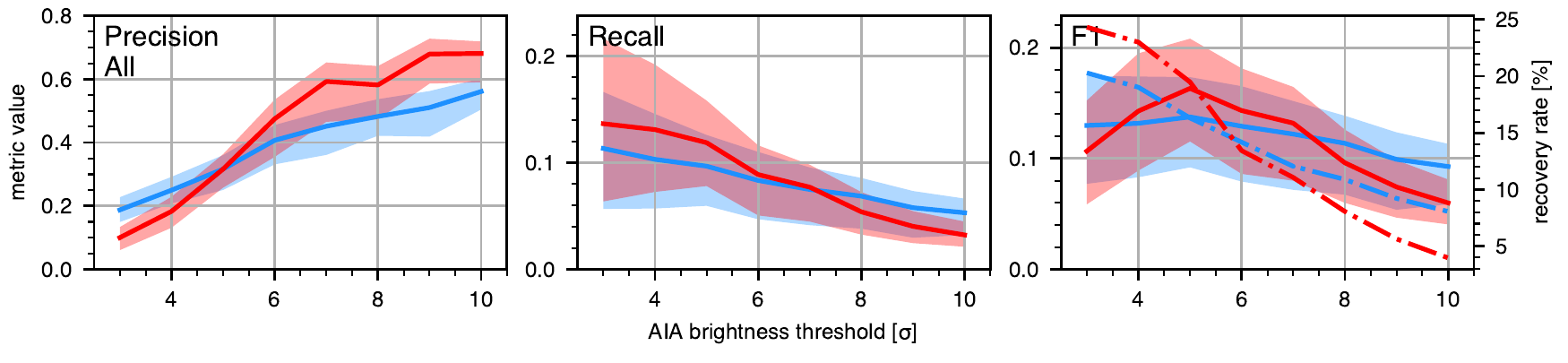}}
  \caption[]{\label{fig:prf_avplot} %
  Average performance metrics with respect to the \Halpha\ detections
  as function of the AIA brightness threshold: precision ({\it
  left\/}), recall ({\it middle\/}) and the \fone-score ({\it
  right\/}).  
  Colour coding as in Fig.~\ref{fig:prf_plots_all}.
  The dash-dotted curves in the third panel (scale at right) specify
  the recovery rates, \ie\ the percentage of \Halpha\ detections with
  true-positive AIA detections, for the parameter combinations giving
  maximum \fone\ per threshold (peak of the \fone\ shading).
  }
\end{figure*}

\paragraph{Performance for  all \Halpha\ detections.}
Figures~\ref{fig:prf_plots_all} and \ref{fig:prf_avplot} summarise the
AIA detection precision, recall and \fone-score as function of imposed
AIA brightness threshold for the individual data sets and averaged over
all data sets, respectively.
In each panel the two AIA passbands are distinguished with colour
coding of the mean curves and of the spread that results from applying
the different additional area and lifetime constraints in our grid of
parameters.
Fig.~\ref{fig:prf_plots_all} shows remarkable variations in curve
behaviour between the different data sets.  
If all would peak at some optimum parameter combination our task would
be easy, but this is not the case.  

FP errors come primarily from pseudo-EBs at low brightness, from FAFs
at high brightness.  
Some FAFS do not pass our stationarity constraint by their fast
apparent motion, but different amounts of remaining FAFs cause
different divergences between the 1600 and 1700\,\AA\ metrics. 
Sets A, B, E, G show no or only few FAFs; sets C, I, and J have many.

Let us first consider the precision P, \ie\ the TP fraction of all AIA
detections (Eq.~\ref{eq:precision}).
The top two rows (panels P-A--P-J) of Figure~\ref{fig:prf_plots_all}
show that it generally increases with brightness threshold.
For most data sets these increases are relatively smooth, but sets A
and H are likely affected by their small-number statistics.
These also show the lowest maximum P (note differences between P axis
scales).
The generally low P values at low threshold come from erroneous
inclusion of pseudo-EBs.
The starting values of the P curves are therefore lowest for fields of
view that contain a large fraction of quiet-Sun (generally 70--80\%
but only around 50\% for sets C, D and I).
To reach high P a high AIA brightness threshold is required, generally
6--8$\sigma$ over the quiet-area average or higher.
A few cases then even reach 100\% (blue curve for set E, red curve for
G). 
In most panels the highest P values are reached with 1700\,\AA\ (red
curve and shading) but sets B and E which contain no FAFs reach
highest P in 1600\,\AA.  

The middle two rows of Fig.~\ref{fig:prf_plots_all} (panels R-A--R-J)
depict the recall R, \ie\ AIA's recovery fraction of all \Halpha\
detections (Eq.~\ref{eq:recall}).
It generally decreases with imposed AIA brightness threshold, as
expected from the predominance of less bright \Halpha\ \EBs\ below the
AIA resolution limit, and it reaches only values below 20\% for most
data sets.  
The behaviour of 1700\,\AA\ is more varied than for 1600\,\AA, showing
steeper decreases in sets A and C and local maxima in B, I and J.
The best performance (highest R reached at the upper border of the
spread envelope) is better for data sets with fewer \Halpha\
detections (A, B, H), which may result from relative paucity of small
and short-lived events that are harder for AIA to replicate. 

The harmonic-mean \fone-score in the bottom two rows of
Fig.~\ref{fig:prf_plots_all} (panels F1-A--F1-J) combines TP maximisation
with FN and FP minimisation (Eq.~\ref{eq:f1}) and so mingles patterns
in the corresponding P and R panels.
Since the latter tend to opposite trends the resulting \fone\ values
are generally poor.
In some cases \fone\ seems almost independent of brightness threshold
regardless of AIA diagnostic (\eg\ sets C and D), while 1600\,\AA\
seems best in A and B and 1700\,\AA\ in I and J but not at the same
brightness threshold.
Roughly, \fone\ peaks at brightness thresholds between 5--7$\sigma$
above the quiet average. 

Figure~\ref{fig:prf_avplot} presents all-data averages of the three
performance metrics weighted by the number of \Halpha-detections per
data set.
The recovery curves at right are similar to but reach higher than the
recall curves in the centre panel, by holding for the best parameter
combination instead of representing the average and because multiple
Halpha\ \EBs\ may contribute to one true-positive AIA detection.
In the last panel 1700\,\AA\ peaks at 5$\sigma$ from balancing the
opposite 1700\,\AA\ slopes in the first two panels. 
At this threshold the average 1700\,\AA\ recall is about 12\%; the
recovery for the best parameter combination including this threshold
reaches over 19\%.
The corresponding precision P (not number-weighted as in the first
panel but the total TP/(TP+FP) for the optimal parameter
combination) is 27\%.

When optimising instead for precision by using a 9$\sigma$
threshold for 1700\,\AA\ then 62\% of the 1700\,\AA\ detections are
\Halpha\ \EBs\ but only 5\% of the \Halpha\ population is recovered.
This choice recovers only the tip of the iceberg. 

\paragraph{Performance for \Halpha\ top fractions only.}
Previous studies have noted, although without statistical analysis,
that typically the largest and brightest \Halpha\ \EBs\ overlap best
with concentrated brightenings in the AIA images, as we find here when
optimising P.
Since the area distribution of the \Halpha\ detections in
Fig.~\ref{fig:halpha_stats} peaks at small values below the
0.36\,\sqasec\ AIA pixel size, it is not surprising that AIA's
recovery is less than 20\% at best and much smaller at higher
thresholds.
We therefore explore the possibility of obtaining higher recovery by
considering only the top of the \Halpha\ population in terms of
lifetime, area, and peak intensity, respectively. 
We also tested a fourth quantity, the total \EB\ intensity obtained by
summing the intensities in all its pixels over its entire lifetime,
but found that this measure (a proxy for total released energy) does
not give significantly better results.

\begin{figure*}[bht]
  \centerline{\includegraphics[width=\textwidth]{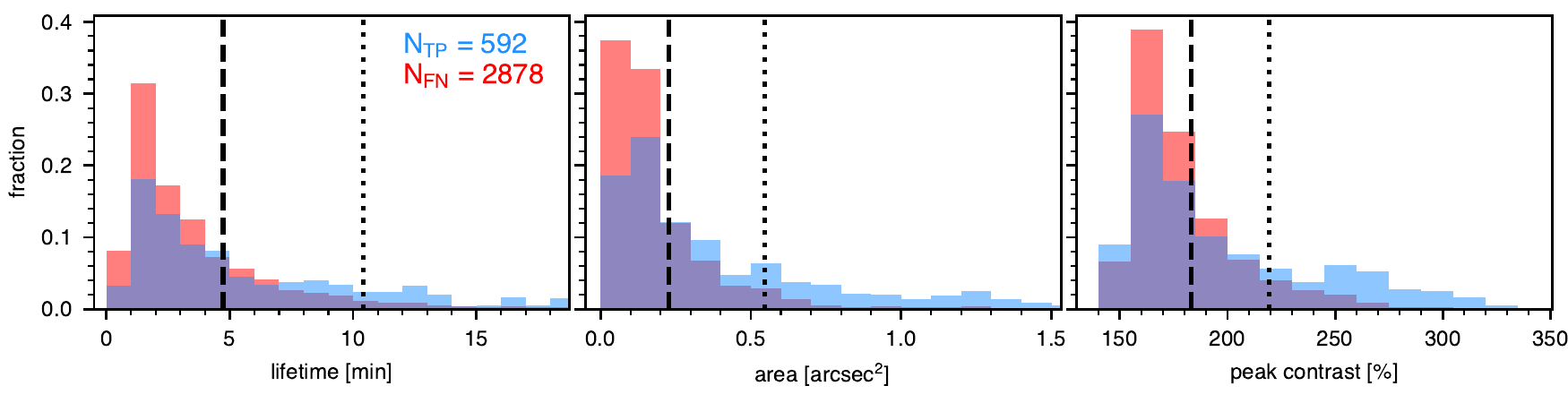}}
  \caption[]{\label{fig:stats_tpfn} %
  Normalised distribution of AIA false negatives FN ({\it red\/}) and
  true positives TP ({\it blue\/}) as function of \Halpha\ \EB\
  lifetime, area, and peak contrast.
  Overlaps show up violet.
  The distributions are summations over the two AIA diagnostics, for
  each using the parameter combination yielding maximum \fone, and are
  normalised by the total numbers specified in the first panel.
  The vertical lines indicate cut-offs for selecting the top 30\%
  ({\it dashed\/}, N\,=\,521) and 10\% ({\it dotted\/}, N\,=\,174) of
  the \Halpha\ \EBs\ for each quantity.
  }
\end{figure*}

\begin{figure*}[bht]
  \centerline{\includegraphics[width=\textwidth]{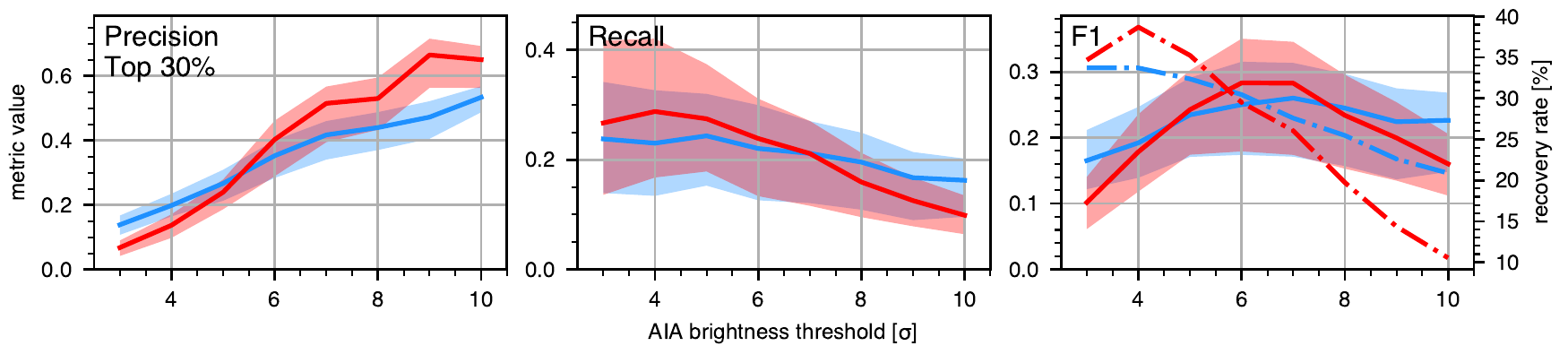}}
  \vspace{-6ex}
  \centerline{\includegraphics[width=\textwidth]{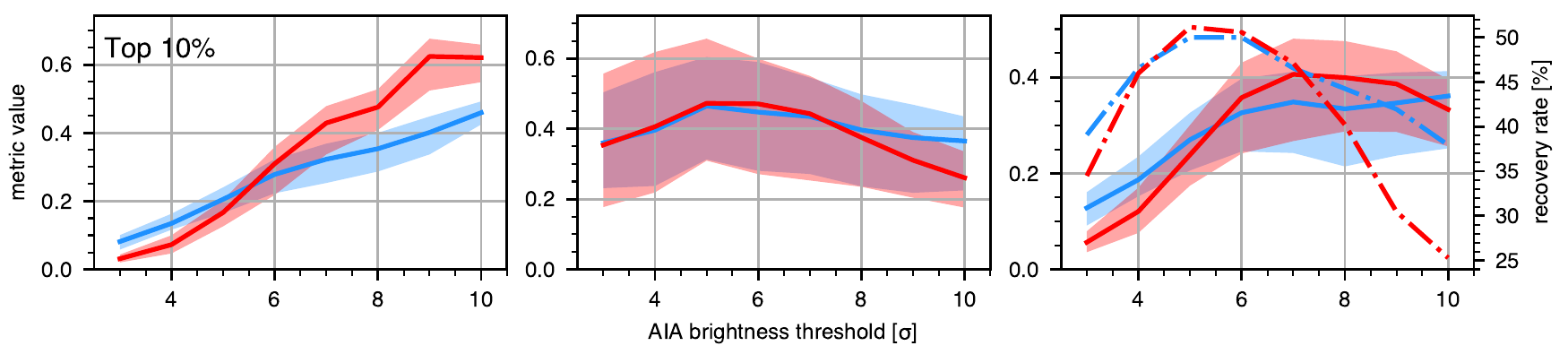}}
  \caption[]{\label{fig:prf_avplot_topsel} %
  Average metrics performance as function of AIA brightness threshold
  for only the top 30\% and 10\% \Halpha\ detections in terms of area.
  Format as for Fig.~\ref{fig:prf_avplot}.
  }
\end{figure*}

Figure~\ref{fig:stats_tpfn} shows AIA FN and TP distributions as
function of the \Halpha\ \EB\ lifetime, area and peak contrast with the
goal to find how to separate FN and TP the best. 
They are summed over the two AIA diagnostics, using for each the
parameter combination that yields the largest \fone.
The FN distributions show that most of the \Halpha\ detections that
AIA missed are small. 
Compared to these the TP distributions have extended high-value tails;
in each panel about half of the TPs fall above the dashed 30\%
boundary while only about 25-30\% of the FNs do so. 
The best separation of the outer TP tail is for area; we therefore
consider the top fractions in this quantity.

Figure~\ref{fig:prf_avplot_topsel} presents these in the format of
Fig.~\ref{fig:prf_avplot}.
The precision trends (first column) behave similar to
Fig.~\ref{fig:prf_avplot} with the values still reaching about 60\%
and decreasing slightly for smaller sub-sample.
AIA 1700\,\AA\ again outperforms 1600\,\AA\ above the 5$\sigma$ threshold.
The recall curves (centre column) now show a more pronounced peak for
1700\,\AA, extending to higher brightness threshold for smaller
sub-sample, whereas those for 1600\,\AA\ are fairly constant with
sub-sample size.  
The recall spread is of order 0.2-0.3, larger than in the centre
panel of Fig.~\ref{fig:prf_avplot}.
The recall values increase significantly for smaller sub-sample size.
AIA 1700\,\AA\ outperforms 1600\,\AA\ only marginally below
brightness threshold 6--7$\sigma$, while 1600\,\AA\ does better above 8$\sigma$.
In the \fone-scores (third column) 1700\,\AA\ again outperforms
1600\,\AA, peaking at 6--7$\sigma$ for the top 30\% sub-sample and at
7--8$\sigma$ for the top 10\% sub-sample.
However, lower brightness thresholds yield better recovery rates
(dot-dashed curves, axis at right), with little difference between 1600\,\AA\
and 1700\,\AA\ below 7$\sigma$.

When optimising for \fone\ by selecting the 1700\,\AA\ 7$\sigma$
threshold both P and R are higher than for the \fone-optimised full
\Halpha\ population but still below 50\%, while the recovery percentage
of all \Halpha\ detections (from the full population) becomes only
5--8\%.
When optimising for P by selecting the 9$\sigma$ threshold the
recovery drops further to 3--5\%. 
The conclusion is that this sub-class selection does not improve the
metrics performances dramatically, while still recovering less than half of the
sub-sample. 

\begin{table}[h]
  \caption{Detection numbers in \Halpha\ and AIA 1700\,\AA.}
\begin{center}
  \begin{tabular}{crrrrrrrrr}
        \hline \hline
  {} & {} & \multicolumn{3}{c}{AIA \fone-optimised} & \multicolumn{3}{c}{AIA P-optimised} \\
  \cline{3-5}\cline{6-8}
  Set & \Halpha\  & TP & FP & FN & TP & FP & FN \\
  \hline
  A & 19  & 5   & 158 & 14  & 1   & 11  & 18 \\
  B & 31  & 11  & 31  & 18  & 1   & 1   & 30 \\ 
  C & 333 & 44  & 53  & 275 & 16  & 5 & 321 \\
  D & 470 & 95  & 182 & 333 & 41  & 32  & 431 \\
  E & 62  & 7   & 42  & 50  & 4   & 1 & 58 \\
  F & 395 & 47  & 133 & 341 & 14  & 1 & 385 \\
  G & 145 & 18  & 10  & 120 & 3   & 0 & 142 \\
  H & 21  & 4   & 21  & 16  & 0   & 1 & 21 \\
  I & 126 & 18  & 13  & 107 & 2   & 2 & 124 \\
  J & 133 & 11  & 44  & 123 & 13  & 4 & 117 \\
  \hline
  Total& 1735 & 260 & 687 & 1397 & 95 & 58 & 1647 \\
        \hline
\end{tabular}
\end{center}
\label{tab:aia_detections}
\end{table}

\begin{figure*}[bht]
  \centerline{\includegraphics[width=\textwidth]{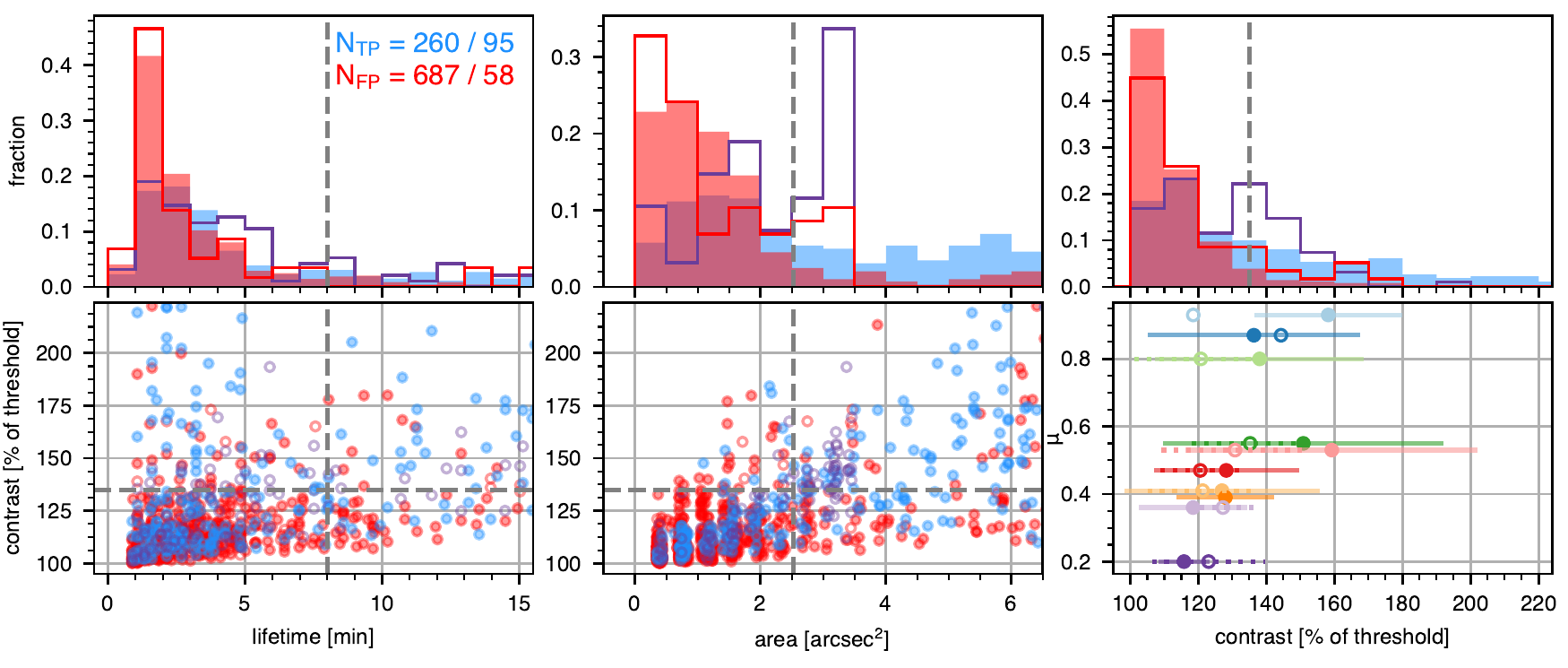}}
  \caption[]{\label{fig:aia_stats} %
  AIA 1700\,\AA\ detection statistics as function of properties.
  Figure layout as for Fig.~\ref{fig:halpha_stats}, except that in all
  but the last panel distinction is made between true positives TP
  (AIA detection with \Halpha\ detection) and false positives FP (AIA
  detection without \Halpha\ detection) rather than data sets.
  The peak contrasts are given as percentage above the intensity threshold
  (quiet-Sun average + 5$\sigma$ for \fone\ maximisation, + 9$\sigma$ for P
  maximisation).
  {\it Upper row\/}: each panel shows the TP ({\it blue / purple\/})
  and FP ({\it red\/}) distribution distributions for the parameter
  combination maximising \fone\ ({\it filled histograms\/}) and P
  ({\it outlined histograms\/}).
  {\it Lower row\/}: as Fig.~\ref{fig:halpha_stats} but the symbols in
  the two scatter plots are colour-coded between TP and FP as in the
  upper row, and with distinction between \fone\ maximisation ({\it
  filled circles\/}) and P maximisation ({\it open circles\/}).
  The last panel shows average peak contrasts for the TP detections with
  the bar lengths showing the rms spreads around the indicated mean
  values and with similar distinction between maximising \fone\ ({\it
  filled circles and solid lines\/}) and maximising P ({\it open
  circles and dotted lines\/}). 
  The total TP and FP numbers are given in the top right of the first
  panel, with slash-separation between \fone\ and P maximisation.
  The vertical and horizontal dashed grey lines in all but the last panel
  indicate the threshold values for sub-selection of the top 10\% longest-lived,
  largest and brightest events.
  }
\end{figure*}

\subsection{Optimal AIA results}
\label{sec:aia_stats}
\paragraph{Optimal  detection parameters.}
Since our trials using only top \Halpha\ detections did not produce
significantly better metrics we define optimal criteria from the
results for all \Halpha\ detections that were summarised in
Fig.~\ref{fig:prf_avplot}. 
On their basis we only employ the 1700\,\AA\ images.
For \fone\ maximisation the criteria are: (1) an intensity threshold
of 5$\sigma$ above the quiet-area average, (2) an area between 1--18
AIA pixels, and (3) a minimum lifetime of 1\,min without upper limit. 
The corresponding metric values are P = 27\%, R = 19\%, \fone\ = 0.23.
If one prefers to instead optimise precision P the first two criteria
become: intensity threshold 9$\sigma$ above the average and area 1--9
AIA pixels. 
The lifetime condition remains the same.
The metrics then become P = 62\%, R = 5\%, \fone\ = 0.09.
Table~\ref{tab:aia_detections} summarises the TP, FP and FN counts
for both \fone and P optimisation.

\paragraph{Properties of AIA-detected \EBs.}
Figure~\ref{fig:aia_stats} shows statistical properties of the AIA
detections resulting from both optimisation recipes, colour-split
between true positives TP and false negatives FP and also splitting
\fone\ maximisation (filled histograms and symbols) and P maximisation
(open histograms and symbols).
The lack of detections above 3.5\,\sqasec\ for the P-maximised area
histograms (outlined) is imposed by the upper area limit of 9 AIA
pixels.
Comparison with the \Halpha-\EB\ statistics in
Fig.~\ref{fig:halpha_stats} shows that the TPs have similar lifetime
(upper left panels in the two figures) and brightness (upper right
panels) distributions, but a different area distribution (upper centre
panels).

The average lifetime and brightness are larger than for \Halpha\
because AIA favours larger features that tend to be brighter and live
longer.
We find lifetimes on the order of 5$\pm$7\,min for both \fone- and P-maximised
detection populations.
The TP and FP lifetime distributions also differ between the two maximisations,
with TPs peaking at lifetimes about 9~min and 6\,min, respectively,
but FPs at about 3--4\,min for both.
Although there is no strong correlation with the intensity contrast
(lower left panel) and many TPs are as faint and short-lived as the
majority of the FPs, the fraction of TPs among the longer-lived events
is larger, with higher contrasts. 
For example, there are no TP detections lasting 15\,min with
brightness contrast below 125\%.

The FP area distribution (upper centre panel) shows a broad peak with
most FP detections below 2\,\sqasec\ regardless of maximisation, 
while part of the TPs has similarly small areas and their majority 
exceeds 2\,\sqasec.
As for the \Halpha\ \EBs, the first two lower panels show higher
correlation between detection area and brightness contrast than for
lifetime. 
In particular, there are no detections smaller than 1\,\sqasec\ with a
contrast above 150\% of the mean, whereas there are several detections
above that contrast that last only 2\,min or less.

The third panel shows that for both \fone\ and P maximisation the FPs
are mostly below 130\% (90\% and 80\% of their numbers) while nearly
half of the TPs are higher.
The panel underneath shows a tendency for the \fone-maximised sample
to have lower contrast closer to the limb, less for the P-maximised
detections (open circles).

\paragraph{Performance for AIA top fractions only.} 
Finally, the precision can be optimised further by recognising that
the false positives FP in Fig.~\ref{fig:aia_stats} cluster at shorter
lifetimes, smaller areas and lower contrasts so that they can be
largely avoided by dropping these samplings altogether.
For using the \fone\ criteria we perform this additional selection by
maintaining only those AIA detections that (1) have a lifetime longer
than 20 AIA frames (8\, min), (2) are larger than 7 AIA pixels
(2.52\,\sqasec) and (3) show peak contrast larger than 135\% of the
5$\sigma$-over-average threshold (9$\sigma$ in the case of the P-maximised
population).
Together these outer-tail selections imply maintaining the top
$\sim$10\% of all AIA detections. 
We found that then the probability that a remaining AIA detection is
an \Halpha\ \EB\ increases from 27\% to 80\%.
Using the same thresholds on P-maximised detections increases the hit
rate even to 87\%. 

\section{Discussion}\label{sec:discussion}

\paragraph{\Halpha\  criteria.}
Our final detection criteria for \Halpha\ \EB\ detection are: (1) a
double core--halo intensity threshold at 145\% and 130\% of the
quiet-Sun average determined from masked HMI data (\ie\ a core of
pixels with brightness of at least 145\%; the halo consists of pixels
of at least 130\% that are adjacent to the core or to other halo
pixels) in either the blue or red wing-average images (constructed
from averaging over $\pm$(0.9--1.1)\,\AA) or both, (2) an area threshold of
0.035\,\sqasec, and (3) a lifetime threshold of approximately 1\,min.

Both intensity thresholds are lower than the ones used by
\citetads{2015ApJ...812...11V} 
from normalisation to quiet-Sun sub-field averaging rather then full
field-of-view averaging.
The same thresholds were employed by
\citetads{2016ApJ...823..110R} 
and they are similar to those of
\citetads{2011ApJ...736...71W} 
and \citetads{1987SoPh..108..227Z}. 

For the area threshold we tested values ranging from 5 to 20 SST
pixels (0.018 to 0.070\,\sqasec) but found through our CRISPEX
inspections that a 5\,pix (0.018\,\sqasec) constraint delivered too
many dubious detections whereas for 15 to 20\,pix too many valid
events were excluded.
We therefore settled on 10\,pix (0.035\,\sqasec) which lies between
the 0.05--0.11\sqasec\ of
\citetads{2016ApJ...823..110R} 
and \citetads{2017GSL.....4...30C}, 
and the $\sim$0.015--0.02\,\sqasec\ value used in \eg\
\citetads{2013ApJ...774...32V}, 
\citetads{2015ApJ...798...19N}, 
\citeads{2015ApJ...812...11V}. 

In our \Halpha\ data the main culprits causing missing or wrong \EB\
identification are too low cadence and too large or too long seeing
deteriorations. 
In addition, the SST resolution sets a lower limit to detectable \EB\
area; it may well be that additional photospheric reconnection events
exist that are even smaller and weaker than the tiny QSEBs of
\citetads{2016A&A...592A.100R}, 
but if so these are unlikely to be picked up at any other optical
telescope nor with AIA (which does not show QSEBs in its ultraviolet
images).

\paragraph{\Halpha\ detections.}
Both the \Halpha-detection lifetime range (majority between
1--15\,min) and its average (roughly 3\,min) compare well with earlier
high-resolution SST studies (cf.~Fig.~\ref{fig:halpha_stats}).  
\citetads{2013ApJ...774...32V} 
reported an average of 3.5--4\,min with 75\% of the \EB\ detections
having lifetimes shorter than 5\,min,
\citetads{2015ApJ...798...19N} 
found lifetimes between 3--20\,min with 7\,min average and
\citetads{2016ApJ...823..110R} 
noted lifetimes ranging 0.5--14\,min peaking around 1\,min.
The typical areas found here (majority between 0.035--0.4\,\sqasec,
average 0.14\,\sqasec) are rather small; both
\citetads{2013ApJ...774...32V} 
and \citetads{2015ApJ...798...19N} 
found averages about 0.2--0.3\,\sqasec.
Earlier \citetads{1987SoPh..108..227Z} 
had found 0.6\,\sqasec.
\citetads{2016ApJ...823..110R} 
reported much larger areas but actually found very similar values with
the majority in the range of 0.06--0.21\,\sqasec\ (private
communication).

We find positive correlation between lifetime and peak intensity
contrast and a stronger correlation between detection area and
intensity contrast (cf.\ first two lower panels of
Fig.~\ref{fig:halpha_stats}), similar to \eg\
\citetads{2015ApJ...798...19N}, 
\citetads{2017A&A...598A..33L}, 
\citetads{2017GSL.....4...30C}. 

\paragraph{AIA 1700\,\AA\ detections.}
Only few statistics exist in the literature regarding \EB-related
detections in mid-ultraviolet continua.
\citetads{2013ApJ...774...32V} 
reported 1.1--1.3\,\sqasec\ average for features detected using a
5$\sigma$-above-mean threshold in AIA 1700\,\AA\ (without area
constraint other than a 0.36\,\sqasec\ lower limit), somewhat lower
than our average of 1.94$\pm$1.75\,\sqasec\ for the total population
including both true and false positives.
\citetads{2007A&A...473..279P} 
found the majority of \EB-related brightenings identified in similar
1600\,\AA\ images from the \TRACE\ (TRACE)
to have lifetimes between 1.5 and 7\,min, with an average at 3.5\,min.
Our true-positive population shows a much higher average of just over
8\,min, but also a large spread with median lifetime only 3.8\,min.

The recent statistical study by
\citetads{2017GSL.....4...30C} 
used a 3.5$\sigma$ above average brightness threshold.
Although they did not report feature sizes, they did note that above a
lifetime of 20\,min AIA-detected \EBs\ dominated their population of
AIA 1700\,\AA\ detections.
In our case, this tipping point lies at only 5.5\,min, with respectively
68\% and 79\% of the AIA detections being true positives when
considering lifetimes above 10 and 20\,min, but this difference may be
explained by the positive correlation between areas and lifetime and
noting that \citetads{2017GSL.....4...30C} 
used a 4 AIA-pixel lower limit.
While not specifically targeting \EBs,
\citetads{2017ApJ...836...63T} 
used a 5$\sigma$ threshold in 1700\,\AA\ to select events for
comparison with \CaIIH\ bright points in an emerging flux region; the
authors argued (and we agree) that many of those were likely \EBs.

\paragraph{Viewing angle effects.}
\EB\ detection sensitivity to viewing angle may be expected. 
On the one hand, \EBs\ are easier recognised towards the limb through
higher contrast (last panel of Fig.~\ref{fig:halpha_stats}) and
because their projected area increases (\eg\ data set A versus E). 
On the other hand, foreshortening at smaller $\mu$ reduces the
projected size of active regions so that more quiet Sun comes into
view (as in sets H and J in Fig.~\ref{fig:fovs}).

We find no clear trend with viewing angle (data set order) in the
lifetimes and maximum detection areas in Table~\ref{tab:halpha_stats}. 
There is a hint of increasing average peak contrast at more limb-ward
viewing (last panel of Fig.~\ref{fig:halpha_stats}), but the standard
deviations are too large to make this significant.
Variations in the inherent activity and evolutionary stage of the
observed target may be more important.
For instance, data set C exhibits an excessively high detection rate,
but this target was part of a highly complex active region with
increasing flux emergence during the time of our observations.
Similarly, sets H and A (sampling the same active region on 27 June
and 2 July 2010) had similar detection rates even though the viewing
angle differed over 0.5.

\paragraph{How suitable are AIA 1700\,\AA\ images to detect \EBs?} 
Ideally, our efforts would have produced a recipe that recovers
\Halpha\ \EBs\ one-to-one from AIA data.
However, as demonstrated in
Figs.~\ref{fig:prf_plots_all}--\ref{fig:prf_avplot} it is not possible
to exclude false detections when optimising \fone\ since neither
precision nor recall then reach unity; optimising for precision alone
does better in that quantity but recovers fewer \Halpha-\EBs.
The result above is that \fone\ optimisation reaches only 27\%
precision (AIA detections that correspond to \Halpha\ detections) and
19\% recall (\Halpha\ detections recovered by AIA), and that these
percentages become 62\% and 5\% when optimising precision
(Sect.~\ref{sec:aia_stats}). 
High recovery was only reached in the subsequent top-10\% AIA
selection.

This lack of one-to-one correspondence has been noted before.
Previous studies found ultraviolet recoveries of \Halpha\ \EBs\ over
50\%
(\citeads{2000ApJ...544L.157Q}, 
\citeads{2002ApJ...575..506G}, 
higher than our results but from \Halpha\ observations with worse
angular resolution.
When we discard \Halpha\ detections below 0.64\,\sqasec\ (the pixel
size in the second study) we obtain a recovery of 66\% (\cf\
Figs.~\ref{fig:stats_tpfn} and \ref{fig:prf_avplot_topsel}).

The recent study by \citeads{2017GSL.....4...30C} 
found precision 53\% and recall 51\%.
Applying their area thresholds of 0.11\,\sqasec\ (three times ours) for
\Halpha\ and 1.44\,\sqasec\ (four times ours) for AIA gives precision
44\% and recall 22\% in our results, but we suspect that the
first threshold was below the effective resolution. 

Moreover, false detections are inevitable.
They are partly explained by the ten-fold resolution difference
between the SST and AIA since the majority of the maximum \Halpha\
\EB\ areas in Fig.~\ref{fig:halpha_stats} is smaller than one AIA
pixel of 0.36\,\sqasec\ (Fig.~\ref{fig:halpha_stats}).
A large class of potential false detections consists of pseudo-EBs
marking magnetic concentrations in quiescent network. 

In addition, there is no reason \emph{per se} to presume that all
\Halpha-observed \EBs\ have counterparts in the ultraviolet.
Several observational studies (%
\citeads{2015ApJ...812...11V}, 
\citeads{2015ApJ...810...38K}, 
\citeads{2016ApJ...824...96T}, 
\citeads{2017A&A...598A..33L}) 
have demonstrated that while the \EB\ and \UVB\ populations overlap,
they do not do so entirely; recent simulation results suggest that the
non-overlaps are at higher reconnection height
(\citeads{2017ApJ...839...22H}). 
Hence, false positives may correspond to \UVBs\ without \EB\
counterpart (\citeads{2015ApJ...812...11V}). 
While regrettable from the perspective of identifying pure \EBs, these
may also serve to trace low-atmosphere reconnection and provide early
warning of emerging flux.
Our final suggestion to consider only the top 10\% fraction of AIA
detections, giving 80\% \EB\ recall excluding all pseudo-EBs, likely
includes these in the remaining 20\% and so may well deliver such
candidates additionally. 

\section{Conclusion}  \label{sec:conclusion}
We implemented \EB\ detection recipes for both imaging spectroscopy in
\Halpha\ and mid-ultraviolet imaging with AIA that improve on earlier
versions.
Key improvements are to consider the \Halpha\ $\pm$(0.9--1.1)\,\AA wings separately and the
definition of viewing-angle and data-set independent quiet-Sun-passing
masks to define brightness thresholds.

We thus detected 1735 \EBs\ in high-quality \Halpha\ observations with
the SST of active regions at ten locations that together span
centre-to-limb viewing and active-region variation. 
This is the first study sampling many \EBs\ from multiple active
regions.
The large variations in Figs.~\ref{fig:halpha_stats} and
\ref{fig:prf_plots_all} show that using only a few in a single
observation to address their visibility in different diagnostics and
their role in the energy and mass budget of the outer atmosphere, as
was done in a number of studies including recent ones, may yield skewed
results and should be done with great care.

We inventoried the corresponding appearance and detectability in
simultaneous AIA 1600 and 1700\,\AA\ images.
With a completeness analysis applying a detection-parameter grid to
these we derived optimal detection criteria to either recover the
largest fraction of \Halpha-\EBs\ while minimising false detections,
or to maximise the number of AIA-detections that are \Halpha-\EBs.
Whether to prioritise the one or the other is a choice that depends on
the purpose of the study.

Overall, detection in AIA 1700\,\AA\ yields the best results. 
Our recommended detection criteria for this diagnostic and the first
choice in prioritising are:
\begin{enumerate} \itemsep=1ex
\item minimum brightness threshold of 5$\sigma$ above the local
quiet-Sun average obtained with masks derived from HMI data;
  \item area limit to between 1 and 18 AIA pixels (0.4--6.5 \sqasec);
  \item minimum lifetime threshold of 1\,min.
\end{enumerate}
These parameter choices should recover about 20\% of the \Halpha-\EB\
population, while ensuring that nearly 30\% of the AIA detections is
indeed an \Halpha-\EB.
Optimising instead for detection precision by using 9$\sigma$
brightness threshold and 1--9\,pix area constraint makes over 60\% of
the AIA events \Halpha-detected \EBs, but at the cost of recovering
only about 5\% of all \Halpha\ \EBs.

Further restriction to only the top 10\% fraction of all AIA
detections that result from the three criteria above can be done by
using additional combined thresholds for lifetime (20 AIA frames),
area (7 AIA pixels) and peak contrast (above 135\% of the 5$\sigma$
value). 
This reaches over 80\% probability that each remaining AIA detection was an
\Halpha-observable \EB.

\EB\ detection in the AIA mid-UV images is thus feasible and can
recover a significant fraction of \Halpha\ \EBs, although full
recovery of the complete \Halpha-\EB\ population that is detectable at
the high resolution and quality of the SST cannot be achieved with the
low-resolution AIA images. 
A fortiori, the smaller but still Ellerman-like QSEB reconnection
events are not observed in AIA's ultraviolet passbands
(\citeads{2016A&A...592A.100R}). 
However, the top 10\% AIA selection furnishes a secure way of finding
the more important ones
and while the recall is then low, some applications (\eg\ early detection of
flux emergence or of active region formation) may not require this but still
benefit from the high precision in AIA detection of \EBs.
In addition, most of the then remaining 20\% false positive detections are
probably \UVB\ candidates without \EB\ counterpart but of interest in their own
right as marking higher-up reconnection.

Altogether, our recipe opens the entire AIA database for performing
continuous, full-disk detection and tracking of low-atmosphere
reconnection events and thereby of flux emergence and magnetic active
region evolution in the past or at present. 
For example, such monitoring may provide valuable input in flare
forecasting.

\begin{acknowledgements}
Our research was supported under the CHROMOBS grant by the Knut and Alice
Wallenberg Foundation, as well as by the ERC under the European Union's Seventh
Framework Programme (FP7/2007-2013)\,/\,ERC grant agreement nr.~291058.
This research was also supported by the Research Council of Norway,
project number 250810, and through its Centres of Excellence scheme,
project number 262622.
The Swedish 1-m Solar Telescope is operated on the island of La Palma
by the Institute for Solar Physics of Stockholm University in the
Spanish Observatorio del Roque de los Muchachos of the Instituto de
Astrof\'isica de Canarias. 
We are grateful to Eamon Scullion for kindly providing the \Halpha\ data of June 20, 2012.
We also want to thank Patrick Antolin, Mats Carlsson, Jaime de la Cruz
Rodr\'iguez, Ainar Drews, Viggo Hansteen, Shahin Jafarzedeh, Torben Leifsen, Ada
Ortiz, Tiago Pereira and Mikolaj Szydlarski for contributing to the other SST
observations used here.
This work benefited from discussions at the meetings ``Solar UV bursts
- a new insight to magnetic reconnection'' at the International Space
Science Institute (ISSI) in Bern.  
We made much use of NASA's Astrophysics Data System Bibliographic Services.
We also acknowledge the community effort to develop open-source
packages used here: \tt{numpy} (\url{numpy.org}), \tt{matplotlib}
(\url{matplotlib.org}), \tt{sunpy} (\url{sunpy.org}).
\end{acknowledgements}

\bibliographystyle{aa}
\bibliography{journals,rjrfiles,adsfiles,eb-aia} 


\end{document}